\documentclass[10pt,twocolumn,letterpaper]{article}

\usepackage{wacv}                           

\usepackage{fontawesome5}
\usepackage{array}
\usepackage{multirow}
\usepackage{algorithm}
\usepackage{algpseudocode}
\usepackage{tikz}
\usetikzlibrary{arrows.meta,positioning}

\renewcommand{\textfraction}{0.07}

\renewcommand{\dbltopfraction}{0.92}
\renewcommand{\dblfloatpagefraction}{0.85}

\newcommand{\rdn}[1]{{\scriptsize(#1\%)}}

\definecolor{wacvblue}{rgb}{0.21,0.49,0.74}
\usepackage[pagebackref,breaklinks,colorlinks,allcolors=wacvblue]{hyperref}

\def\wacvPaperID{1683} 
\def\confName{WACV}
\def\confYear{2027}

\title{Reconstruction-Aware Cryo-EM Particle Picking}

\author{Riku Itsuji\textsuperscript{1,2}\quad
Yuanhao Wang\textsuperscript{3}\quad
Xingjian Li\textsuperscript{3}\quad
Seonghui Min\textsuperscript{3}\quad
Hideo Saito\textsuperscript{1,2}\quad
Min Xu\textsuperscript{3}\\[4pt]
\textsuperscript{1}Keio University, Yokohama, Japan\quad
\textsuperscript{2}Keio AI Research Center, Yokohama, Japan\\
\textsuperscript{3}Carnegie Mellon University, Pittsburgh, PA, USA\\[7pt]
{\small
\href{https://riku359.github.io/recon-aware-pick-page/}{\faGlobe\,\ Project Page}\quad
\href{https://github.com/riku359/ReconAwarePick}{\faGithub\,\ GitHub}\quad
\href{https://huggingface.co/datasets/rikrikrik/recon-aware-pick-data}{\faDatabase\,\ Dataset}\quad
\href{https://huggingface.co/rikrikrik/recon-aware-pick-weights}{\faBrain\,\ Weights}}
}

\begin{document}
\maketitle
\AddToShipoutPicture*{\AtTextLowerLeft{\raisebox{-16pt}{\small Preprint.}}}
\begin{abstract}
Cryo-electron microscopy (cryo-EM) determines the structures of proteins and macromolecular assemblies at near-atomic resolution, and the final 3D reconstruction depends on extracting a clean particle stack from noisy micrographs. This extraction decomposes into three sub-tasks, namely particle picking, contamination removal, and 2D class selection. Each of them, however, is trained and evaluated in isolation, and none is optimized for the reconstruction. We instead integrate the three sub-tasks into a single pipeline posed against downstream reconstruction quality. We instantiate the pipeline with a state-of-the-art component for each sub-task, CryoTransformer picking permissively, MicrographCleaner masking contamination, and CryoSift selecting 2D classes by a continuous quality score, and close the loop with a fine-tuning step that returns the surviving particles to the picker. The pipeline achieves a better 3D resolution than every picker we compare. We also show that the best 2D F1 is not the best resolution, so particle selection is better treated as one reconstruction-aware pipeline judged by the map it delivers.
\end{abstract}

\section{Introduction}
\label{sec:intro}

Cryo-electron microscopy (cryo-EM) has become a central technique for determining the structures of proteins and macromolecular assemblies at near-atomic resolution \cite{kuhlbrandt2014,nakane2020,yip2020}. Most of these structures are solved by single-particle analysis (SPA), in which a large number of noisy 2D projections of many identical copies of the same molecule are combined into a single 3D density map \cite{cheng2015primer}. The workflow first images the vitrified specimen into 2D micrographs, each containing projections of many particles frozen in random orientations. The particles located and extracted from the micrographs form the stack that 3D reconstruction consumes. This work addresses that middle stage, whose goal is to deliver a particle stack that supports a high-quality reconstruction.

Two properties of the data make this stage difficult. First, electron dose must be kept low to limit radiation damage, so the contrast between a particle and background is small. Second, high-contrast contamination such as ice crystals and carbon edges is present in the same micrographs and can be picked as false positives.

To obtain a usable particle set under these conditions, current practice decomposes the stage into three sub-tasks. \emph{Particle picking} localizes candidates in each micrograph, as done by crYOLO \cite{cryolo2019}, Topaz \cite{topaz2019}, CryoTransformer \cite{cryotransformer2024}, and CryoSegNet \cite{cryosegnet2024}. \emph{Contamination removal} masks these contaminated regions and discards the candidates that fall inside them, as in MicrographCleaner \cite{micrographcleaner2020}. \emph{2D classification} groups the extracted particles into classes of similar projections, and a selection step such as CryoSift \cite{cryosift2025} rejects the non-particle classes.

Each of these sub-tasks performs well on its own, but optimizing them separately does not necessarily give a good 3D reconstruction. Particle picking, for instance, is usually scored by precision, recall, and F1 against expert-annotated coordinates, most often on CryoPPP \cite{cryoppp2023}, and the trade-off against false positives risks overlooking rare views. In a pipeline where contamination removal and 2D classification already remove outliers, a picker can tolerate false positives so that rare views are not missed.

The purified set is also a source of supervision for the picker. 2D classification is particularly informative here, because it groups similar views into classes and lets the decision be made on a class average rather than on a single noisy particle. The surviving particles can be mapped back to their micrographs and reused as pseudo-labels, which adapts the picker to the target data without manual annotation.

Only a few existing works tackle this combined setting. TranSPHIRE \cite{transphire2020} and McSweeney \etal \cite{mcsweeney2020} close the loop by retraining the picker with feedback from 2D classification. Both select the classes with a heuristic or binary decision that cannot express the quality differences between classes. Neither removes high-contrast contamination, which can survive to form coherent false-positive 2D classes.

We propose \emph{reconstruction-aware particle picking}, a formulation in which detection and purification are designed jointly against downstream reconstruction quality rather than separately against 2D metrics. We instantiate it as a pipeline that assembles a state-of-the-art model for each sub-task and feeds the purified set back to the picker. The high-recall picker CryoTransformer \cite{cryotransformer2024} proposes candidates, MicrographCleaner \cite{micrographcleaner2020} removes contamination, and CryoSift \cite{cryosift2025} scores the 2D class averages of the survivors on a continuous scale from $1.0$ to $5.0$. Finally, the purified set fine-tunes the picker, which re-picks every micrograph before the two purification stages run again.

We evaluate the pipeline on the four EMPIAR entries that form CryoTransformer's independent test set (EMPIAR-10081, 10093, 10345, and 10532), processing the full micrograph set of each entry rather than only the 300 micrographs annotated by CryoPPP. We judge it not by 2D detection metrics but by the final reconstruction resolution obtained from the stack it delivers. The pipeline improves on CryoTransformer, the picker it is built on, by 0.63 to 3.57\,\AA{} across the four entries, and gives the best resolution of any picker we compare.

\vspace{0.5em}
\noindent Our contributions are summarized as follows:
\begin{itemize}
    \itemsep0em
    \item \textbf{Reconstruction-aware formulation.} We formulate particle picking, contamination removal, and 2D class selection as a single selection problem posed against downstream reconstruction, and evaluate it by reconstruction resolution rather than by 2D F1 against incomplete annotations; reconstruction is used for system-level evaluation, not as a direct training signal.
    \item \textbf{A purification pipeline that outperforms existing pickers.} We combine a high-recall picker with contamination masking and 2D class selection, and close the loop by re-training the picker on the purified particle stack. The pipeline gives the best reconstruction of any picker we compare.
    \item \textbf{The best F1 is not the best reconstruction.} Across the same pickers and entries, 2D detection scores do not reliably order the reconstructions. Agreement with 2D annotations is therefore only a partial guide, and we judge a selection by the map it delivers.
\end{itemize}

\section{Related Work}
\label{sec:related}

\noindent\textbf{Particle picking.} Early pickers match reference projections as templates \cite{findem2004,gempicker2013} or detect particle-sized blobs with hand-designed filters \cite{dogpicker2009,applepicker2018,kltpicker2020}. Classical machine learning followed, with classifiers over hand-crafted features \cite{mallick2004,autopicker2014,autocryopicker2019,supercryoempicker2019}. Deep learning now dominates the task. CNN-based pickers came first and remain the most used \cite{deeppicker2016,deepem2017,cryolo2019,topaz2019,drpnet2021}. Among them, crYOLO \cite{cryolo2019} adapts a single-shot object detector but tends to overlook true particles, while Topaz \cite{topaz2019} learns from positive-unlabeled data and admits many false positives in return \cite{cryomae2025,upicker2024}. Transformer-based detectors came next \cite{transpicker2021,cryotransformer2024,gtpick2025}, of which CryoTransformer \cite{cryotransformer2024} reaches high recall on the CryoPPP benchmark \cite{cryoppp2023} at the cost of false positives and duplicate picks \cite{cryomae2025,upicker2024,vmpicker2026}. Other work casts picking as segmentation \cite{pixer2019,warp2019,cassper2021,urdnet2022,crisp2025} and more recently builds on the Segment Anything Model \cite{cryosegnet2024,cryopromptseg2026,cryosip2026,vmpicker2026}, where CryoSegNet \cite{cryosegnet2024} pairs a denoising U-Net with SAM and reports large gains over earlier pickers. Few-shot pickers \cite{cryomae2025,cryofsl2026,cryoanomaly2026} and semi-supervised, self-supervised, and synthetic-data alternatives \cite{upicker2024,cryoemmae2025,parseek2026} reduce the annotation cost. Vanaja Pandi \etal \cite{contrastiveclustering2025} make the detector robust to noisy and incomplete annotations with a contrastive loss and clustering. Across all of these approaches, however, picking is formulated as an isolated 2D detection task, optimized and ranked by precision, recall, and F1 against incomplete annotations.

\noindent\textbf{Contamination removal.} Contamination is removed either at the level of whole micrographs or at the level of regions within one. At the micrograph level, a classifier labels each micrograph as good or bad \cite{micassess2020,miffi2024,prismpyp2026,tripletfewshot2025}. At the region level, classical methods delineate contaminated areas through filtering and geometric fitting or through variational segmentation \cite{emhp2017,asocem2022}. Learned segmentation models instead predict a per-pixel contamination mask \cite{micrographcleaner2020,icefinder2025,qwencryomarker2026,cryosparcjunkdetector}, among which MicrographCleaner \cite{micrographcleaner2020} is trained on the most diverse data, 539 expert-segmented micrographs drawn from 16 EMPIAR entries. A further line uses contamination as a supervision signal for the picker itself, as artifact classes in the detection objective \cite{warp2019,pixer2019,cassper2021,urdnet2022,eman2tools2018,drpnet2021}, as a veto on proposals during pseudo-label generation \cite{upicker2024}, or as a negative-supervision term \cite{cryoanomaly2026}. However, no picker has yet used the mask to purify the particle set that is fed back to the picker.

\noindent\textbf{2D classification.} 2D classification groups extracted particle images into classes of similar projections and averages them into denoised class images, as done by the maximum-likelihood frameworks of RELION \cite{relion2012} and CryoSPARC \cite{cryosparc2017} and by a range of alternative clustering formulations \cite{isac2012,primecluster2016,ackmeans2016,re2dc2022,dcaekmeans2023,simcryocluster2024}. Selecting the good output of classification was long manual and subjective, motivating automatic selection at two levels. At the particle-image level, individual particles are ranked or rejected \cite{zhousorting2020,ppcasorting2023,cryosparcautoclustering,cryoief2026,fraser2021,pointclouddeform2026}. At the class-average level, 2DAssess \cite{micassess2020} assigns four quality categories and Cinderella \cite{sphirecinderella} a binary good or bad label, whereas CryoSift \cite{cryosift2025} scores class quality on a continuous scale \cite{relionclassranker2021,cryosparcautoselect2d}. More fundamentally, selection terminates the pipeline in all of these methods, and the labels it produces are never returned to the picker.

\noindent\textbf{Feedback-driven pipelines.} Only a few systems close the loop from downstream curation back to the picker. McSweeney \etal \cite{mcsweeney2020} select 2D classes by a heuristic rule and iteratively retrain a CNN picker on the surviving particles. TranSPHIRE \cite{transphire2020} operates on the fly during acquisition. It labels class averages good or bad with Cinderella \cite{sphirecinderella} and retrains crYOLO on the good classes over several feedback rounds. The loops above select with heuristic thresholds or binary labels that cannot express nuanced quality differences \cite{mcsweeney2020,transphire2020}, and neither includes a stage for removing contamination. Our pipeline replaces the binary decision with the continuous class-quality scores of CryoSift \cite{cryosift2025}, and places a contamination mask predicted automatically by MicrographCleaner \cite{micrographcleaner2020} between picking and classification, so that high-contrast candidates are removed before the classes are formed.

\section{Method}
\label{sec:method}

\begin{figure*}[t]
  \centering
  \includegraphics[width=\linewidth]{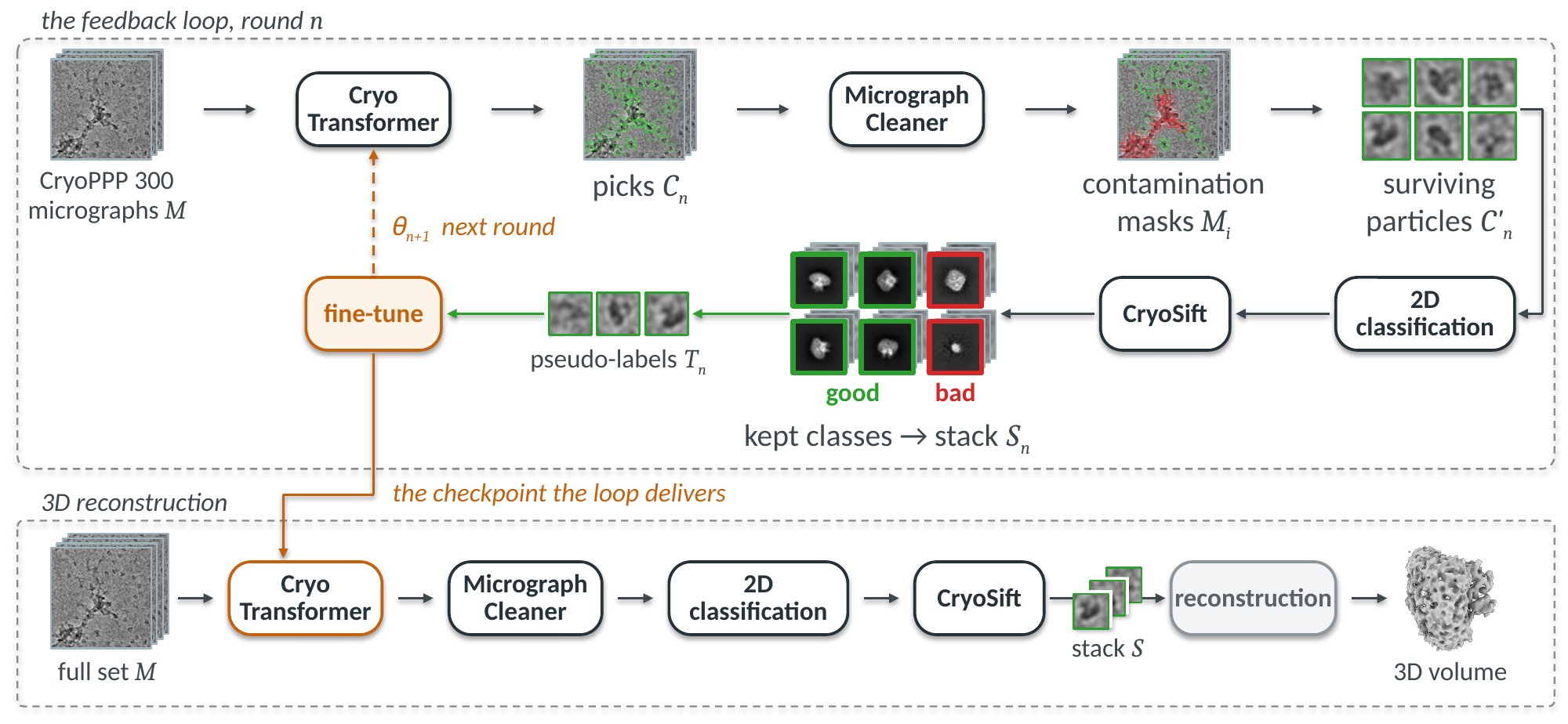}
  \caption{\textbf{Reconstruction-aware particle picking.} Top: one round of the feedback loop. CryoTransformer picks candidates, a MicrographCleaner mask discards the candidates on contamination, and 2D class selection keeps the classes CryoSift scores as good. The surviving particles serve as pseudo-labels for fine-tuning the picker. Bottom: the checkpoint the loop delivers picks the full micrograph set, the same masking and 2D class selection purify the picks, and the surviving stack goes to 3D reconstruction.}
  \label{fig:pipeline}
\end{figure*}

\subsection{Problem formulation}
\label{sec:method:formulation}

\Cref{fig:pipeline} summarizes our pipeline, and Sec.~S1 of the supplementary material draws the same pipeline as the CryoSPARC jobs. Let $\mathcal{M}$ be the micrographs of one dataset and let $f_\theta$ be a picker producing a candidate set $\mathcal{C}=f_\theta(\mathcal{M})$ of particle coordinates. A selection operator $\sigma$ returns a stack $\mathcal{S}=\sigma(\mathcal{C})\subseteq\mathcal{C}$, which is passed to 3D reconstruction. We treat the picker and the selection operator as one system and score it by the reconstruction computed from $\mathcal{S}$, namely its resolution and the number of particles retained (\cref{sec:setup:metrics}).

The selection operator applies two purification stages in turn. Contamination masking removes individual candidates that fall on contaminated regions of the micrograph (\cref{sec:method:cleaner}), and 2D class selection removes entire classes of similar candidates after 2D classification (\cref{sec:method:cryosift}). Both stages can only discard, so no stage recovers a particle the picker never proposed. The pipeline therefore suits a high-recall picker. We close the loop and update the picker on the survivors of $\sigma$ (\cref{sec:method:feedback}).

\subsection{Recall-first candidate generation}
\label{sec:method:pick}

We build on CryoTransformer \cite{cryotransformer2024}, a transformer-based detector that emits a fixed set of scored candidate queries per micrograph. We choose it for its high recall, as reported on the CryoPPP benchmark \cite{cryotransformer2024,cryomae2025,upicker2024}. We run the picker to over-pick, accepting background candidates in exchange for missing few true particles. We keep the operating point of the original implementation at every round and for every dataset. Section~S7 of the supplementary material confirms the recall on our four entries, and Sec.~S2 gives the settings of the operating point.

\subsection{Contamination masking}
\label{sec:method:cleaner}

Contamination such as carbon film, ice, and aggregates carries more contrast than the particles, so candidates taken from it would otherwise reach 2D classification and survive there as false-positive classes. We remove them beforehand with the pretrained model of MicrographCleaner \cite{micrographcleaner2020}, which predicts contamination as a per-pixel probability map. A candidate is discarded when the predicted mask, resized to the full micrograph resolution, reaches a probability of 0.5 at the candidate's center.

\subsection{2D class selection}
\label{sec:method:cryosift}

This stage removes the background classes that over-picking admits, and it decides at the level of a class rather than of a single candidate. The surviving candidates are extracted and grouped by the 2D classification job of CryoSPARC \cite{cryosparc2017} into classes of particles with similar appearance, with the class count left at its default of $K=50$. Each class average is scored by CryoSift \cite{cryosift2025} on a continuous scale from 1.0 (clean particle classes) to 5.0 (non-particle classes). To keep true particles that a single binary cut would take away with the non-particles, we follow the iterative workflow of the CryoSift paper. The best-scoring classes are set aside so that rare views can form classes of their own, the worst are permanently rejected, and the rest are re-classified for a fixed number of cycles. A final classification over the survivors and the set-aside classes then completes the selection. Section~S4 of the supplementary material draws this workflow.

\subsection{Reconstruction-aware feedback}
\label{sec:method:feedback}

\begin{figure*}[t]
  \centering
  \includegraphics[width=0.75\linewidth]{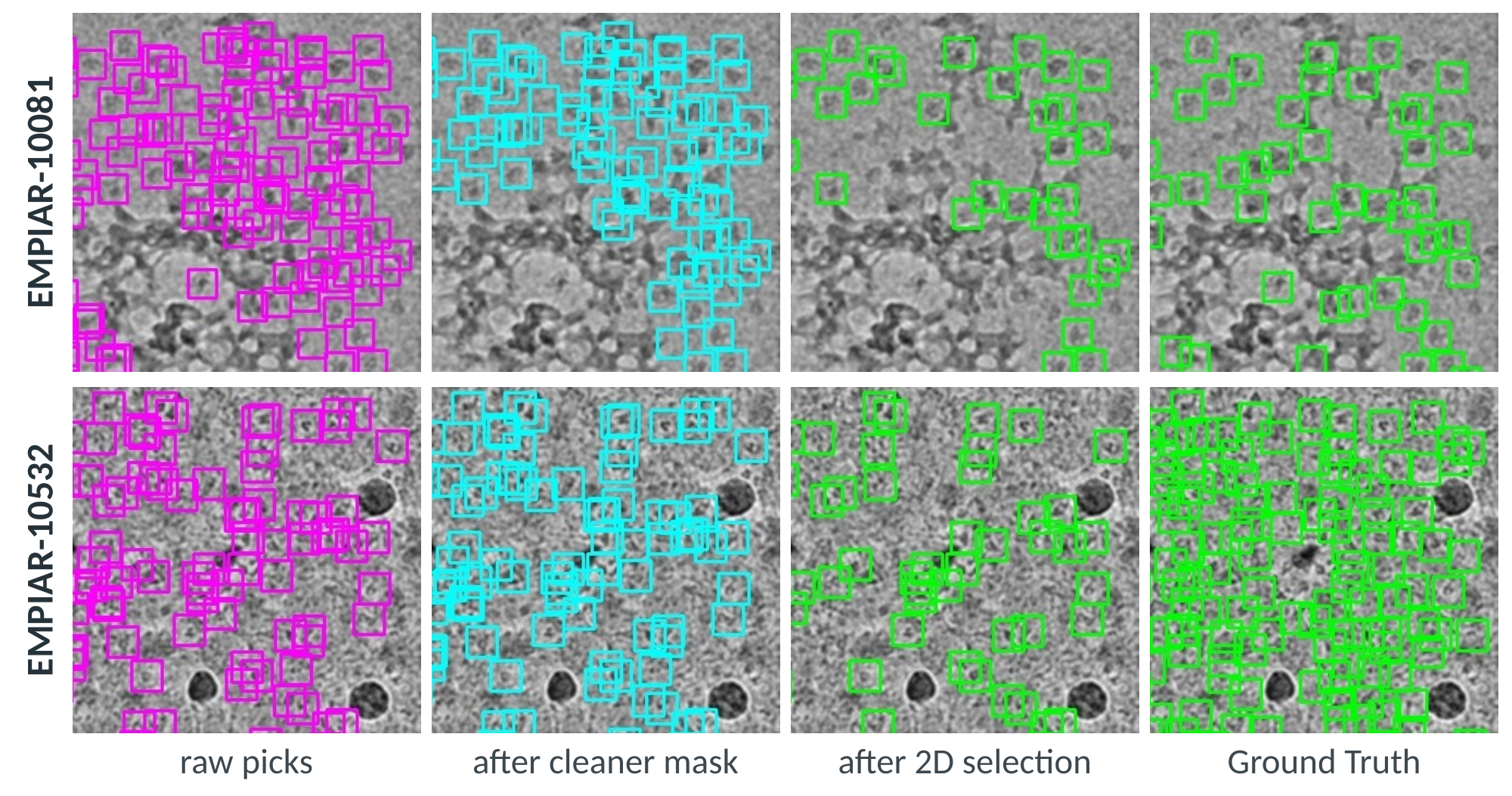}
  \caption{\textbf{What survives each stage of the pipeline.} The picks that remain after
  each stage on one micrograph of EMPIAR-10081 (top) and one of EMPIAR-10532 (bottom) at
  round 1, with the annotated particles repeated on the right. Qualitative counterpart of
  \cref{tab:particles}.}
  \label{fig:pick_fates}
\end{figure*}

The particles that survive contamination masking and 2D class selection serve as pseudo-labels for fine-tuning the picker. Each round maps the survivors back to micrograph coordinates and fine-tunes the current model to reproduce them on micrographs that contain survivors (\cref{fig:pick_fates}), giving the update
\begin{equation}
  \theta_{n+1}=\mathrm{FineTune}\bigl(\theta_{0};\,\mathcal{S}_n\bigr),\qquad
  \mathcal{S}_n=\sigma\bigl(f_{\theta_n}(\mathcal{M})\bigr),
  \label{eq:update}
\end{equation}
where $\theta_n$ are the picker weights after round $n$ and $\theta_0$ is the general CryoTransformer checkpoint. Following TranSPHIRE \cite{transphire2020}, every round restarts the fine-tune from $\theta_0$. The update trains every weight of the picker except the first residual stage of the backbone, which stays frozen.

\section{Experimental Setup}
\label{sec:setup}

\subsection{Datasets}
\label{sec:setup:data}

We evaluate on the four entries of the Electron Microscopy Public Image Archive (EMPIAR) \cite{empiar2016} that form the independent test set of CryoTransformer \cite{cryotransformer2024}, summarized in \cref{tab:datasets}. The CryoPPP dataset \cite{cryoppp2023} provides expert particle annotations for 300 micrographs per entry, which supply the ground truth for the 2D detection metrics and serve as the working set of the feedback loop (\cref{sec:results:loop}). All reconstruction-level results use the full unannotated deposition of each entry, because 300 micrographs are not enough for a stable reconstruction.

\begin{table}[t]
  \centering
  \small
  \caption{\textbf{Datasets.} The four EMPIAR entries of CryoTransformer's test set.
  CryoPPP annotates 300 micrographs per entry (annot.), while the full deposition
  (full) carries none. $d$ is the nominal particle diameter.}
  \label{tab:datasets}
  \setlength{\tabcolsep}{4pt}
  \begin{tabular}{lccccc}
    \toprule
    EMPIAR & Sample & $d$ (px) & psize (\AA) & annot. & full \\
    \midrule
    10081 & HCN1 & 154 & 1.30 & 300 & 997 \\
    10093 & NOMPC & 172 & 1.22 & 300 & 1{,}873 \\
    10345 & integrin $\alpha_{\mathrm{V}}\beta_{8}$ & 149 & 0.673$^{\dagger}$ & 300 & 1{,}644 \\
    10532 & hemagglutinin & 174 & 1.03 & 300 & 1{,}556 \\
    \bottomrule
  \end{tabular}
  \\[2pt]
  {\footnotesize $^{\dagger}$As declared by CryoPPP. Multiply by about two
  when comparing against EMDB.}
\end{table}

\subsection{Metrics}
\label{sec:setup:metrics}

\noindent\textbf{3D reconstruction.} The primary measure is the resolution of the final map, the gold-standard FSC between two independently refined half-maps at the 0.143 criterion \cite{rosenthal2003}. The reconstruction runs three times with different random seeds, and we report the best of the three maps by GSFSC resolution. We also report how many particles reach the reconstruction, and read local-resolution maps as a qualitative check.

\noindent\textbf{2D detection.} As an auxiliary signal, we report precision, recall, and F1 against the expert annotations of CryoPPP \cite{cryoppp2023}. Predicted and annotated centers are matched one-to-one in order of increasing Euclidean distance, and a match within the particle radius $d/2$ (\cref{tab:datasets}) counts as a true positive. Scores are averaged over annotated micrographs and computed with the same code for every picker, independently of any picker-specific confidence score. The same metrics against the per-round pseudo-labels track the feedback loop.

\subsection{Implementation details}
\label{sec:setup:impl}

\noindent\textbf{Base picker checkpoint.} The released CryoTransformer weights carry a training defect that leaves the classification head uninformative. We repair the head and use the repaired checkpoint as $\theta_0$, the starting point of every condition in this paper. The repair is described in Sec.~S2 of the supplementary material.

\noindent\textbf{Fine-tuning.} Each round fine-tunes the picker for 50 epochs, with the hyperparameters listed in Sec.~S2.

\noindent\textbf{Feedback loop.} Each round's teacher set uses 50 micrographs by default, the same budget as TranSPHIRE \cite{transphire2020}, sampled with a fixed per-round seed from the micrographs containing surviving particles and split 40/10 into training and validation. We run three fine-tuning rounds. The loop itself runs no reconstruction, because at its 300-micrograph scale a reconstruction does not resolve the difference between one round and the next. We follow the rounds through the 2D metrics and the number of picked particles instead. To separate the quality of the teacher labels from the design of the loop, we also run one round with the CryoPPP annotations of the same 50 micrographs in place of the surviving picks, holding everything else fixed.

\noindent\textbf{2D class selection.} This stage runs the iterative workflow of \cref{sec:method:cryosift} with the default thresholds of CryoSift \cite{cryosift2025}, whose box-size rule sets the number of cycles to three on all four datasets. The values of the thresholds are given in Sec.~S4 of the supplementary material.

\noindent\textbf{Contamination masking.} Masks are predicted with the released MicrographCleaner weights \cite{micrographcleaner2020}. We replace the post-processing that assembles the windowed predictions of the network into one mask per micrograph, because the released post-processing can spread the mask over clean regions and remove the particles picked there. Section~S3 of the supplementary material describes the replacement.

\label{sec:setup:recon}
\noindent\textbf{Reconstruction protocol.} Every condition is reconstructed with the same CryoSPARC v4.7 \cite{cryosparc2017} job chain. Section~S1 of the supplementary material draws the chain as it runs.

\noindent\textbf{Hardware.} The hardware and the compute time of each stage are given in Sec.~S9 of the supplementary material.

\section{Results}
\label{sec:results}

\subsection{Main results}
\label{sec:results:main}

We compare crYOLO \cite{cryolo2019}, Topaz \cite{topaz2019}, CryoSegNet \cite{cryosegnet2024}, and CryoTransformer \cite{cryotransformer2024} as alternative pickers. The raw picks of each picker are reconstructed with the protocol of \cref{sec:setup:recon}, without the contamination masking and 2D selection of our pipeline. CryoTransformer serves as the picker in all other experiments of this paper.

\begin{figure*}[tp]
  \centering
  \includegraphics[width=\linewidth]{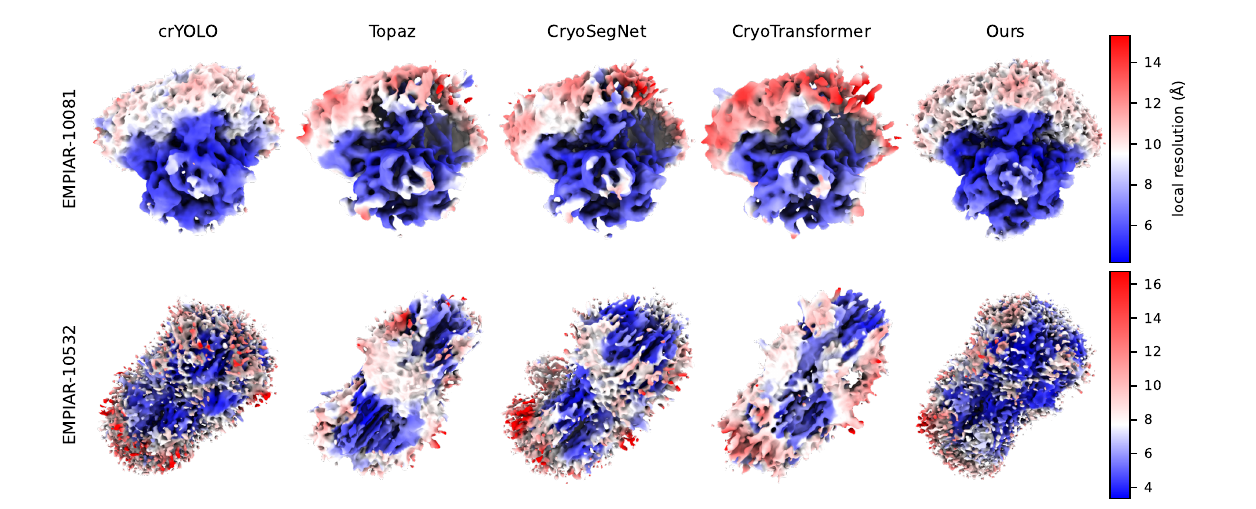}
  \caption{\textbf{3D reconstructions of EMPIAR-10081 and 10532.} The best-of-three map of
  each picker and of our pipeline, colored by the local resolution estimated on it, blue
  better and red worse, on a scale shared within a row. Qualitative counterpart of
  \cref{tab:main_results}.}
  \label{fig:maps}
\end{figure*}

CryoTransformer on its own gives the worst reconstruction in \cref{tab:main_results}. But built on this picker, our pipeline gives the best resolution on all four entries, improving on CryoTransformer by 0.63 to 3.57\,\AA{}. The local resolution map in \cref{fig:maps} also shows that our map resolves the most fine-grained structure of the five. Section~S6 of the supplementary material carries the FSC curve and the viewing-direction distribution of \cref{tab:main_results} and \cref{tab:ablation_res}.

\begin{table}[t]
  \centering
  \small
  \caption{\textbf{Main results.} GSFSC 0.143 resolution (\AA, best of three seeds,
  lower is better) on the full micrograph sets.
  Bold is best and underline second best per dataset.}
  \label{tab:main_results}
  \begin{tabular}{lcccc}
    \toprule
    Method & 10081 & 10093 & 10345 & 10532 \\
    \midrule
    crYOLO          & \underline{4.25} & \underline{4.55} & \underline{3.56} & \underline{3.54} \\
    Topaz           & 4.56 & 6.55 & 3.88 & 3.96 \\
    CryoSegNet      & 4.47 & 6.96 & 3.76 & 3.86 \\
    CryoTransformer & 5.02 & 6.78 & 7.11 & 4.03 \\
    \midrule
    Ours            & \textbf{4.12} & \textbf{4.37} & \textbf{3.54} & \textbf{3.40} \\
    \bottomrule
  \end{tabular}
\end{table}

\subsection{Ablation study}
\label{sec:results:ablation}

We compare four conditions that start from the same CryoTransformer candidate set, \emph{baseline} (raw picks), $+$\emph{mask} (MicrographCleaner contamination masking, \cref{sec:method:cleaner}), $+$\emph{select} (CryoSift iterative 2D class selection, \cref{sec:method:cryosift}), and $+$\emph{both} (masking followed by 2D selection). The \emph{fb} condition instead re-picks the full micrograph set with the round-1 feedback checkpoint (\cref{sec:method:feedback}), and applies the same masking and 2D selection on top. The choice of round 1 follows the loop diagnostics in \cref{sec:results:loop}.

Contamination masking helps on three of the four entries and costs 0.41\,\AA{} on EMPIAR-10093 (\cref{tab:ablation_res}). \cref{tab:mask_removals} classifies the removed candidates against the CryoPPP annotations to analyze this outcome. The mask helps on EMPIAR-10081 and 10345 because almost everything it removes is a non-particle. On EMPIAR-10532, however, 16.6\% of the removals are annotated particles, so the mask purifies this entry at the cost of genuine particles (\cref{sec:discussion:failure}).

\begin{table}[t]
  \centering
  \small
  \caption{\textbf{Mistaken rate of the contamination mask.} Candidates the mask removes on
  the annotated micrographs, split by whether they overlap a CryoPPP-annotated particle
  (\cref{sec:setup:metrics}). The mistaken rate is the share of particles among the
  removals.}
  \label{tab:mask_removals}
  \begin{tabular}{lrrrr}
    \toprule
    Removed & 10081 & 10093 & 10345 & 10532 \\
    \midrule
    particle            &   166 &  4 &    13 &   910 \\
    non-particle        & 4,564 & 36 & 1,923 & 4,564 \\
    \midrule
    mistaken rate (\%)  &   3.5 & 10.0 &  0.7 & 16.6 \\
    \bottomrule
  \end{tabular}
\end{table}

2D selection gives the largest improvement of any stage, with or without the contamination mask in front of it (\cref{tab:ablation_res}). Applied to the raw picks it improves every entry, from 0.56\,\AA{} on EMPIAR-10532 to 3.56\,\AA{} on EMPIAR-10345, and adding the mask in front of it improves three of the four entries a little further, by at most 0.09\,\AA. The candidate counts in \cref{tab:particles} show where this improvement comes from. Contamination masking removes at most 5.6\% of the raw picks, while masking and 2D selection together remove more than 50\% of them on every entry. That count also includes the candidates lost at extraction and 2D classification, and almost all of the removal is the work of 2D selection, the stage that also moves the resolution.

Re-picking with the feedback checkpoint on top of both stages improves three of the four entries again, by at most 0.15\,\AA, and costs 0.04\,\AA{} on the fourth.

\begin{table}[t]
  \centering
  \small
  \caption{\textbf{Ablation study.} GSFSC 0.143 resolution (\AA, best of three seeds) as
  the stages between picking and reconstruction are toggled. fb re-picks every micrograph
  with the round-1 feedback checkpoint and corresponds to Ours in \cref{tab:main_results}.
  Bold is best and underline second best per dataset.}
  \label{tab:ablation_res}
  \begin{tabular}{cccrrrr}
    \toprule
    mask & select & fb & 10081 & 10093 & 10345 & 10532 \\
    \midrule
               &            &            & 5.02 & 6.78 & 7.11 & 4.03 \\
    \checkmark &            &            & 4.74 & 7.19 & 6.96 & 3.98 \\
               & \checkmark &            & 4.15 & 4.61 & \underline{3.55} & 3.47 \\
    \checkmark & \checkmark &            & \textbf{4.08} & \underline{4.52} & 3.59 & \underline{3.43} \\
    \checkmark & \checkmark & \checkmark & \underline{4.12} & \textbf{4.37} & \textbf{3.54} & \textbf{3.40} \\
    \bottomrule
  \end{tabular}
\end{table}

\begin{table}[t]
  \centering
  \small
  \caption{\textbf{Particles retained through the pipeline.} Raw picks and what is left
  with the mask alone and with the mask and 2D selection together. The value in
  parentheses is the cumulative reduction (\%) from the raw picks of the same entry.}
  \label{tab:particles}
  \setlength{\tabcolsep}{3pt}
  \begin{tabular*}{\linewidth}{@{\extracolsep{\fill}}ccrrrr@{}}
    \toprule
    mask & select & 10081 & 10093 & 10345 & 10532 \\
    \midrule
               &            & 259,335 & 754,434 & 494,061 & 604,430 \\
    \midrule
    \multirow{2}{*}{\checkmark} &            & 244,924 & 753,440 & 484,649 & 576,473 \\
               &            & \rdn{-5.6} & \rdn{-0.1} & \rdn{-1.9} & \rdn{-4.6} \\
    \midrule
    \multirow{2}{*}{\checkmark} & \multirow{2}{*}{\checkmark} & 126,181 & 301,589 & 29,214 & 178,722 \\
               &            & \rdn{-51.3} & \rdn{-60.0} & \rdn{-94.1} & \rdn{-70.4} \\
    \bottomrule
  \end{tabular*}
\end{table}

\subsection{Feedback loop}
\label{sec:results:loop}

\Cref{tab:loop_rounds} follows the loop round by round, scoring each round's checkpoint against the CryoPPP annotations and counting the remaining picks after purification. On every entry except EMPIAR-10345, both the particle counts and the macro F1 scores settle at round 1. Rounds 2 and 3 stay within 0.013 macro F1 of it. We therefore report round 1 throughout this paper.

The direction of the round-1 change differs across the datasets. On EMPIAR-10093 and 10532, the precision increases as intended. A larger proportion of the picks survives purification, meaning the picker learns to avoid candidates that 2D selection would discard. Conversely, on EMPIAR-10081, the picker over-picks, increasing recall at the expense of precision. On EMPIAR-10345, the loop instead degrades the picker. The F1 of every round stays below that of the base checkpoint, because the pseudo-labels 2D selection leaves on this entry are sparse (\cref{sec:discussion:failure}).

We also compare the pseudo-labels against a perfect teacher to establish the upper bound of this approach. Replacing them with the CryoPPP annotations of the same 50 micrographs results in almost the same final resolution (\cref{tab:teacher_quality}). What bounds the loop here is therefore not the quality of its labels. \Cref{sec:discussion:subtractive} takes up what does.

\begin{table}[t]
  \centering
  \small
  \setlength{\tabcolsep}{4pt}
  \caption{\textbf{Per-round loop diagnostics} on the 300 annotated micrographs.
  Round 0 picks with the base checkpoint, round $n$ with the checkpoint trained on the
  teacher labels of round $n-1$. The after purify.\ block counts the particles left after
  masking and 2D selection. P / R / F1 are macro scores against the CryoPPP annotations,
  best of the four rounds in bold.}
  \label{tab:loop_rounds}
  \footnotesize
  \begin{tabular*}{\linewidth}{@{\extracolsep{\fill}}lrlrr@{}}
    \toprule
          & \multicolumn{2}{c}{picks} & \multicolumn{2}{c}{after purify.} \\
    \cmidrule(lr){2-3} \cmidrule(lr){4-5}
    round & count & P / R / F1 & count & share \\
    \midrule
    \multicolumn{5}{@{}l}{\emph{EMPIAR-10081}} \\
    0 &  65,385 & \textbf{0.530} / 0.919 / \textbf{0.655} & 33,844 & 51.8\% \\
    1 &  81,787 & 0.470 / 0.970 / 0.610 & 37,234 & 45.5\% \\
    2 &  80,114 & 0.481 / \textbf{0.971} / 0.619 & 37,935 & 47.4\% \\
    3 &  83,272 & 0.465 / \textbf{0.971} / 0.606 & 37,261 & 44.7\% \\
    \midrule
    \multicolumn{5}{@{}l}{\emph{EMPIAR-10093}} \\
    0 & 114,176 & 0.370 / \textbf{0.737} / 0.491 & 39,021 & 34.2\% \\
    1 &  93,962 & 0.429 / 0.703 / \textbf{0.532} & 47,007 & 50.0\% \\
    2 &  90,145 & \textbf{0.434} / 0.683 / 0.530 & 42,490 & 47.1\% \\
    3 &  89,453 & 0.433 / 0.675 / 0.526 & 40,761 & 45.6\% \\
    \midrule
    \multicolumn{5}{@{}l}{\emph{EMPIAR-10345}} \\
    0 &  71,074 & \textbf{0.232} / \textbf{0.947} / \textbf{0.354} &  4,458 &  6.3\% \\
    1 &  67,992 & 0.194 / 0.821 / 0.298 &  3,611 &  5.3\% \\
    2 &  76,690 & 0.159 / 0.726 / 0.248 &  3,965 &  5.2\% \\
    3 &  74,838 & 0.164 / 0.745 / 0.257 &  3,349 &  4.5\% \\
    \midrule
    \multicolumn{5}{@{}l}{\emph{EMPIAR-10532}} \\
    0 & 108,453 & 0.459 / \textbf{0.580} / \textbf{0.498} & 32,213 & 29.7\% \\
    1 &  81,983 & \textbf{0.517} / 0.487 / 0.490 & 38,421 & 46.9\% \\
    2 &  80,747 & 0.509 / 0.469 / 0.477 & 40,148 & 49.7\% \\
    3 &  81,548 & 0.506 / 0.474 / 0.479 & 36,281 & 44.5\% \\
    \bottomrule
  \end{tabular*}
\end{table}

\begin{table}[t]
  \centering
  \small
  \caption{\textbf{Pseudo-labels against a perfect teacher.} GSFSC 0.143 resolution
  (\AA, best of three seeds) of the full micrograph set after one round of the loop, with
  the teacher taken from the surviving picks or from the CryoPPP annotations of the same
  50 micrographs.}
  \label{tab:teacher_quality}
  \setlength{\tabcolsep}{4pt}
  \begin{tabular*}{\linewidth}{@{\extracolsep{\fill}}lrrrr@{}}
    \toprule
    teacher labels & 10081 & 10093 & 10345 & 10532 \\
    \midrule
    surviving picks (Ours)     & 4.12 & 4.37 & 3.54 & 3.40 \\
    CryoPPP annotations & 4.07 & 4.49 & 3.54 & 3.43 \\
    \bottomrule
  \end{tabular*}
\end{table}

\section{Discussion}
\label{sec:discussion}


\subsection{The best 2D F1 is not the best reconstruction}
\label{sec:discussion:f1}

\begin{figure}[t]
  \centering
  \includegraphics[width=\linewidth]{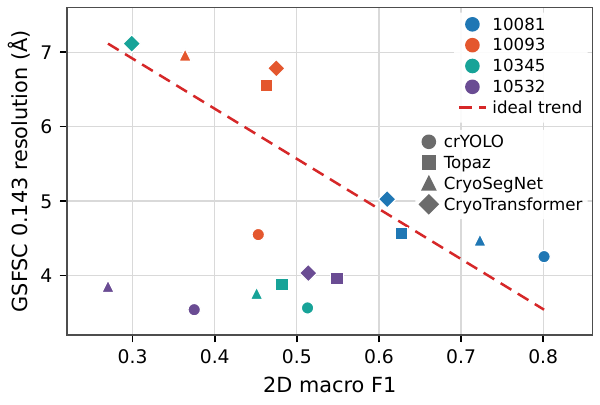}
  \caption{\textbf{2D F1 does not order the pickers the way reconstruction does.} The macro
  F1 of the four base pickers on the 300 annotated micrographs against the GSFSC 0.143
  resolution of \cref{tab:main_results}, with marker shape the picker and color the
  dataset. The red dashed line is a guide to the eye for reading orderings within one
  color, not distances across colors.}
  \label{fig:f1_vs_res}
\end{figure}

\noindent\textbf{F1 and resolution across the base pickers.} \Cref{fig:f1_vs_res} plots the macro F1 of the four base pickers against the resolution the same stacks reach in \cref{tab:main_results}. The F1 values behind the plot are tabulated in Sec.~S7 of the supplementary material. The pickers reach their highest F1 on EMPIAR-10081, and it is the only entry where the F1 and resolution rankings agree. On the other three entries, where every picker has a low F1, the two rankings differ, and on EMPIAR-10532 they are close to reversed. The 2D score therefore predicts the resolution ranking only where the pickers reach a high F1.

\noindent\textbf{The same selection on another picker.} CryoSegNet shares the CryoPPP training data of CryoTransformer \cite{cryoppp2023} and has the higher precision on every entry and the higher F1 on EMPIAR-10081 and 10345, as Sec.~S7 tabulates. On its own it has the better resolution on three of the four entries (\cref{tab:main_results}). We apply the same contamination masking and 2D class selection to its candidate set (\cref{tab:cryosegnet_purified}). After the same selection, the CryoTransformer set is ahead on all four, and so is our full pipeline. The picker with the lower F1 wins once the later stages are included, because its high-recall candidates still hold particles that CryoSegNet has already dropped. A picker is therefore worth optimizing as part of the pipeline rather than by its 2D F1 alone.

\begin{table}[t]
  \centering
  \small
  \caption{\textbf{The same selection applied to CryoSegNet and CryoTransformer} (GSFSC 0.143,
  \AA, best of three seeds, lower is better) on the full sets. The $+$ mask $+$ select rows
  apply our contamination masking and 2D selection to each picker's candidate set. Bold is
  best and underline second best per dataset.}
  \label{tab:cryosegnet_purified}
  \setlength{\tabcolsep}{2pt}
  \begin{tabular}{lcccc}
    \toprule
    Condition & 10081 & 10093 & 10345 & 10532 \\
    \midrule
    CryoSegNet                          & 4.47 & 6.96 & 3.76 & 3.86 \\
    CryoSegNet $+$ mask $+$ select      & 4.14 & 5.71 & 3.78 & 3.74 \\
    CryoTransformer $+$ mask $+$ select & \textbf{4.08} & \underline{4.52} & \underline{3.59} & \underline{3.43} \\
    \midrule
    Ours                                & \underline{4.12} & \textbf{4.37} & \textbf{3.54} & \textbf{3.40} \\
    \bottomrule
  \end{tabular}
\end{table}

\subsection{Failure case analysis}
\label{sec:discussion:failure}

\noindent\textbf{2D selection on EMPIAR-10345.} CryoSift discards classes by an absolute threshold on a score whose scale is not comparable across datasets. On EMPIAR-10345 the whole distribution has shifted past that threshold, so the first classification discards 44 of the 50 classes and 88.6\% of the particles (\cref{fig:cryosift_scores}). Section~S5 of the supplementary material visualizes what this first cycle keeps and discards. The surviving classes are also the teacher, so every later round discards more than 90\% of the picks and stays below the F1 of the base checkpoint (\cref{tab:loop_rounds}). Future work should handle such a shift of the score distribution.

\begin{figure}[t]
  \centering
  \includegraphics[width=\linewidth]{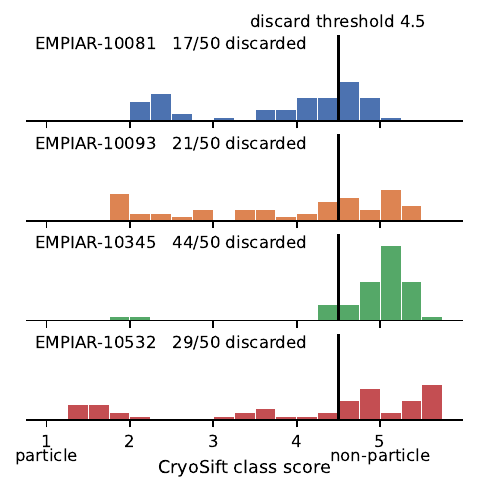}
  \caption{\textbf{The score scale of 2D selection is not comparable across datasets.}
  CryoSift scores of the 50 classes of each entry's first 2D classification, with the
  absolute discard threshold of 4.5 marked.}
  \label{fig:cryosift_scores}
\end{figure}

\begin{figure}[t]
  \centering
  \includegraphics[width=\linewidth]{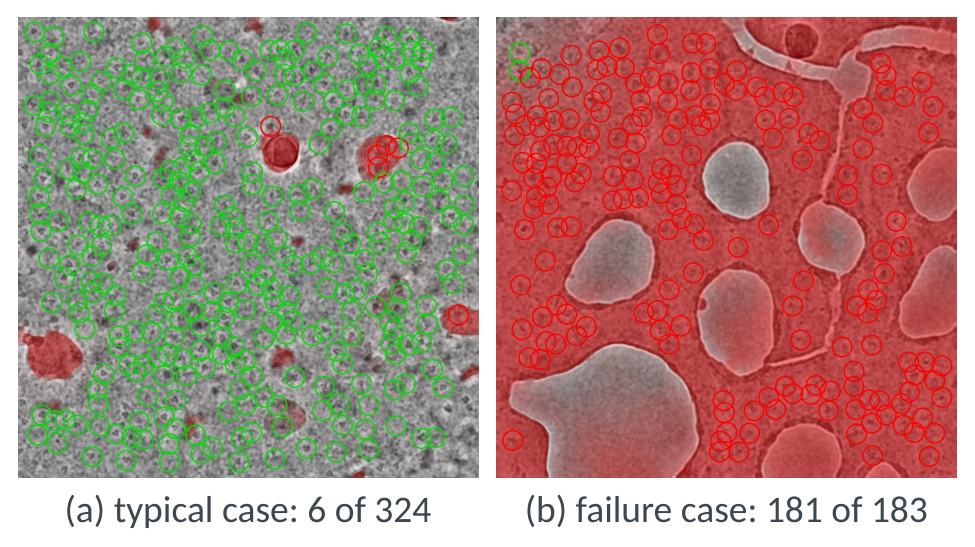}
  \caption{\textbf{The contamination mask fails on part of EMPIAR-10532.} The predicted
  mask in red over the denoised micrograph, with annotated particles green where kept and
  red where removed. Each panel counts the annotated particles it loses.}
  \label{fig:cleaner_failure}
\end{figure}

\noindent\textbf{Contamination masking on EMPIAR-10532.} This is the entry where the mask removes annotated particles at the highest rate (\cref{tab:mask_removals}), and the removals concentrate on a few micrographs. Four of the 300 annotated micrographs lose more than half of their matched particles. On them the mask covers the particle field. The contamination is left as bright holes in it, not as the dark blobs the mask covers elsewhere (\cref{fig:cleaner_failure}). EMPIAR-10532 is not included in the training data of the released MicrographCleaner weights, so the pipeline inherits the training domain of the pretrained model. Future work should tackle domain shifts of this kind, such as contamination that is brighter than the particles rather than darker.

\subsection{Limitations of the subtractive design}
\label{sec:discussion:subtractive}

As \cref{tab:teacher_quality} shows, the loop does not substantially improve the resolution even when the CryoPPP annotations replace the pseudo-labels in the re-training. What bounds it is therefore the direction of the decision rather than the quality of its labels. Both purification stages can only discard candidates (\cref{sec:method:formulation}), and the loop teaches the picker to stop proposing the candidates that those stages would remove anyway. In practice, an expert reads the 2D classes and picks again to collect more particles of an under-represented view, with the class selection of CryoSPARC \cite{cryosparcselect2d} and a Topaz picker \cite{topaz2019} retrained on the kept classes. Future work should build the loop around such an additive step, which we expect to improve the reconstruction further.

\section{Conclusion}
\label{sec:conclusion}

We formulated particle picking, contamination removal, and 2D class selection as a single selection problem posed against the reconstruction rather than against 2D agreement with incomplete annotations. We instantiated it with CryoTransformer, MicrographCleaner, CryoSift, and a fine-tuning step on the picker. The pipeline gives the best resolution of any picker we compare on all four EMPIAR entries, and the 2D ranking of those pickers does not reproduce the ranking of their reconstructions. The problem is therefore better treated at the level of the pipeline, by purifying a high-recall picker, than by optimizing the 2D F1 of the picker alone.


\iftoggle{wacvfinal}{
\noindent\textbf{Acknowledgments.} This work was supported in part by U.S. NSF grants DBI-2238093, DBI-2422619, IIS-2211597, and MCB-2205148. We used Claude (Anthropic) for grammar correction and for brainstorming ideas during writing. The authors are responsible for the entire content of the paper.
}{}

{
    \small
    \bibliographystyle{ieeenat_fullname}
    \bibliography{references}
}

\clearpage
\renewcommand{\dbltopfraction}{0.98}
\renewcommand{\textfraction}{0.02}
\renewcommand{\dblfloatpagefraction}{0.85}
\setcounter{section}{0}
\setcounter{table}{0}
\setcounter{figure}{0}
\renewcommand{\thesection}{S\arabic{section}}
\renewcommand{\thetable}{S\arabic{table}}
\renewcommand{\thefigure}{S\arabic{figure}}
\twocolumn[{%
  \centering
  {\Large\bf Reconstruction-Aware Cryo-EM Particle Picking\\[3pt]
   Supplementary Material\par}
  \vspace{1.4em}
}]

\noindent This supplementary material follows the order of the main paper. \Cref{sec:supp:protocol} draws the pipeline as the CryoSPARC jobs that run it. \Cref{sec:supp:training,sec:supp:maskpostproc,sec:supp:cryosift} give the implementation details of the picker, the contamination mask, and the 2D class selection, and \cref{sec:supp:selection} shows how that selection fails on EMPIAR-10345. \Cref{sec:supp:diagnostics} collects the FSC curves and the viewing-direction distributions behind the tables of the main paper, and \cref{sec:supp:f1scores} tabulates the 2D detection scores of the base pickers. \Cref{sec:supp:limitations} lists two limitations of the evaluation, and \cref{sec:supp:cost} reports the hardware and the compute time.

\section{Details of the pipeline}
\label{sec:supp:protocol}

\begin{figure*}[p]
  \centering
  \definecolor{pcaside}{RGB}{44,160,44}
  \definecolor{pcpool}{RGB}{44,116,200}
  \definecolor{pcdrop}{RGB}{206,62,40}
  \begin{tikzpicture}[
    x=1cm, y=1cm,
    >={Latex[length=1.6mm,width=1.4mm]},
    pan/.style={inner sep=0pt, outer sep=0pt, draw=black!45, line width=0.25pt,
                anchor=west},
    car/.style={inner sep=0pt, outer sep=0pt, anchor=center},
    lab/.style={font=\scriptsize, align=center, inner sep=1pt, anchor=north,
                text width=2.6cm},
    wlab/.style={lab, text width=3.4cm},
    job/.style={draw, rounded corners=2pt, align=center, inner sep=3pt, anchor=west,
                font=\scriptsize, text width=2.10cm, minimum height=9mm},
    proc/.style={job, text width=1.95cm, minimum height=8mm},
    sub/.style={draw=black!45, dotted, rounded corners=3pt},
    blk/.style={draw=black!55, dashed, rounded corners=3pt},
    tag/.style={anchor=west, font=\scriptsize\itshape, inner sep=1pt, fill=white},
    ar/.style={->, semithick, black!70},
    ln/.style={semithick, black!70},
  ]

  \node[proc] (act) at (0.10,0)  {CryoTransformer};
  \node[pan]  (a1)  at (2.62,0)  {\includegraphics[width=2.00cm]{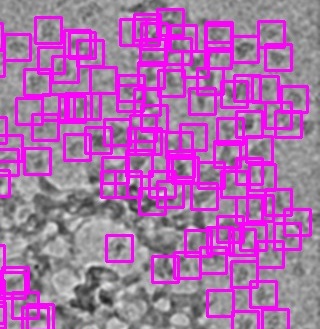}};
  \node[proc] (amc) at (4.98,0)  {MicrographCleaner};
  \node[pan]  (a2)  at (7.50,0)  {\includegraphics[width=2.00cm]{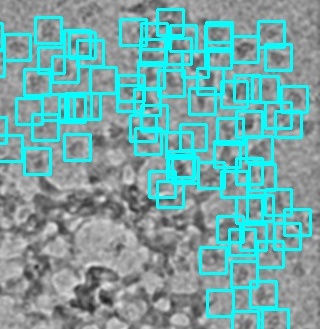}};
  \node[proc] (aex) at (9.86,0)  {Extract};
  \node[pan]  (a3)  at (12.38,0) {\includegraphics[width=1.70cm]{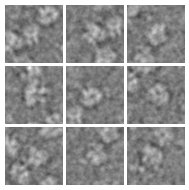}};
  \node[job]  (aft) at (14.72,0) {fine-tune on\\pseudo-labels\\$\theta_{n+1}$};

  \node[lab] at (3.62,-1.18) {picks};
  \node[lab] at (8.50,-1.18) {surviving picks};
  \node[lab] at (13.23,-1.18) {extracted particles};

  \draw[ar] (act.east) -- (a1.west);
  \draw[ar] (a1.east) -- (amc.west);
  \draw[ar] (amc.east) -- (a2.west);
  \draw[ar] (a2.east) -- (aex.west);
  \draw[ar] (aex.east) -- (a3.west);

  \node[pan] (ac0) at (0.00,-3.70)  {\includegraphics[width=3.40cm]{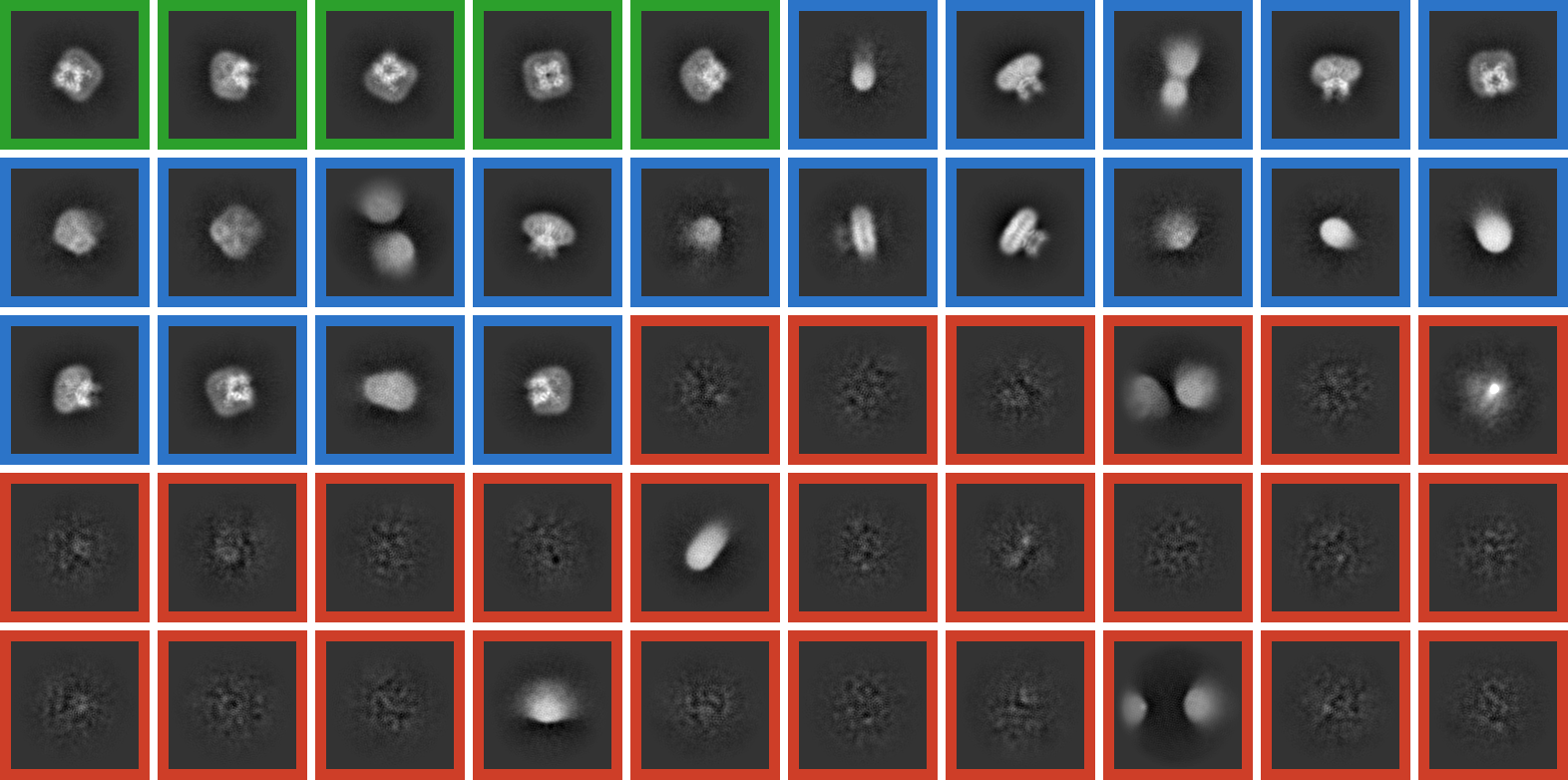}};
  \node[pan] (ac1) at (4.50,-3.70)  {\includegraphics[width=3.40cm]{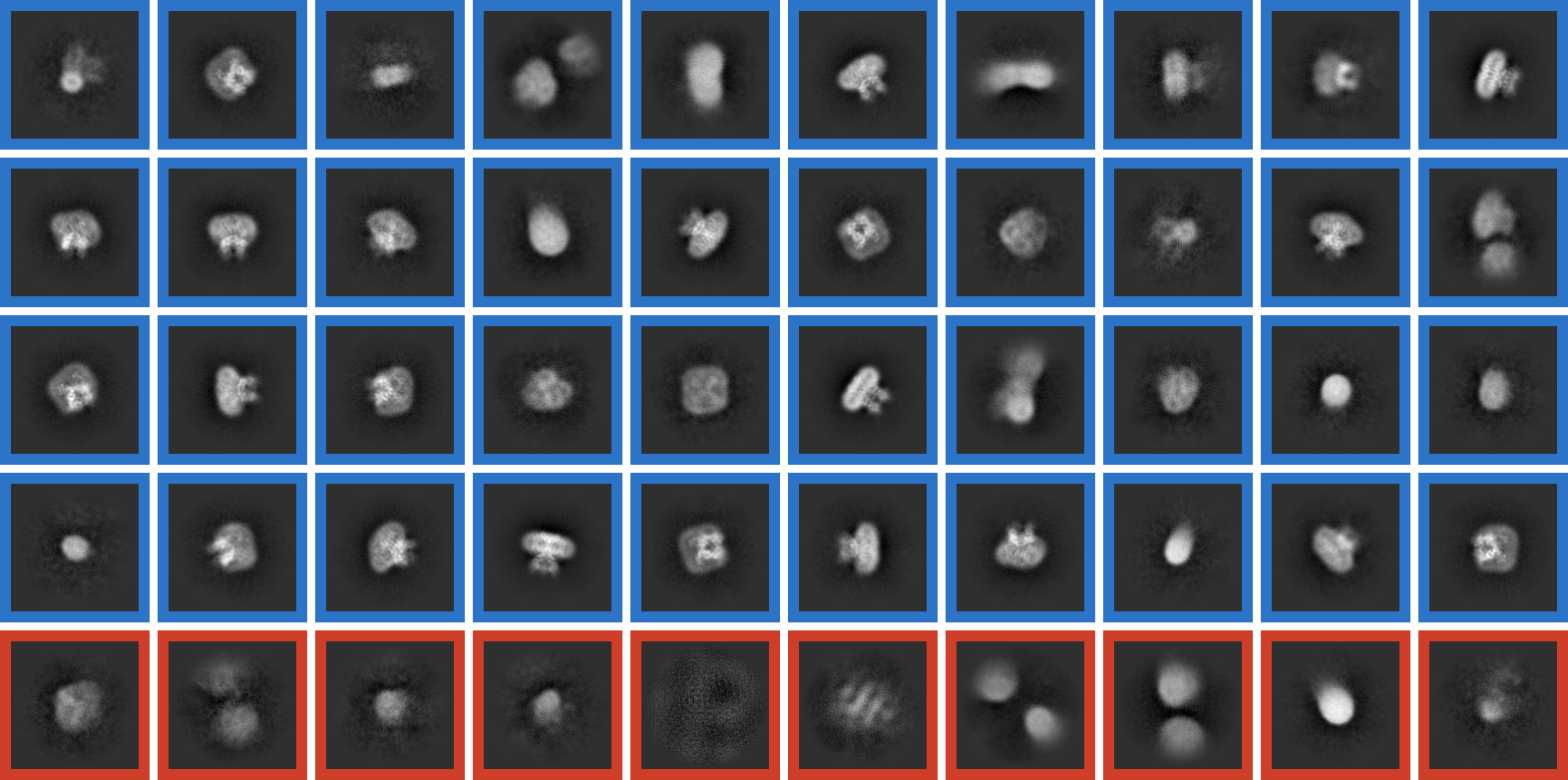}};
  \node[pan] (ac2) at (9.00,-3.70)  {\includegraphics[width=3.40cm]{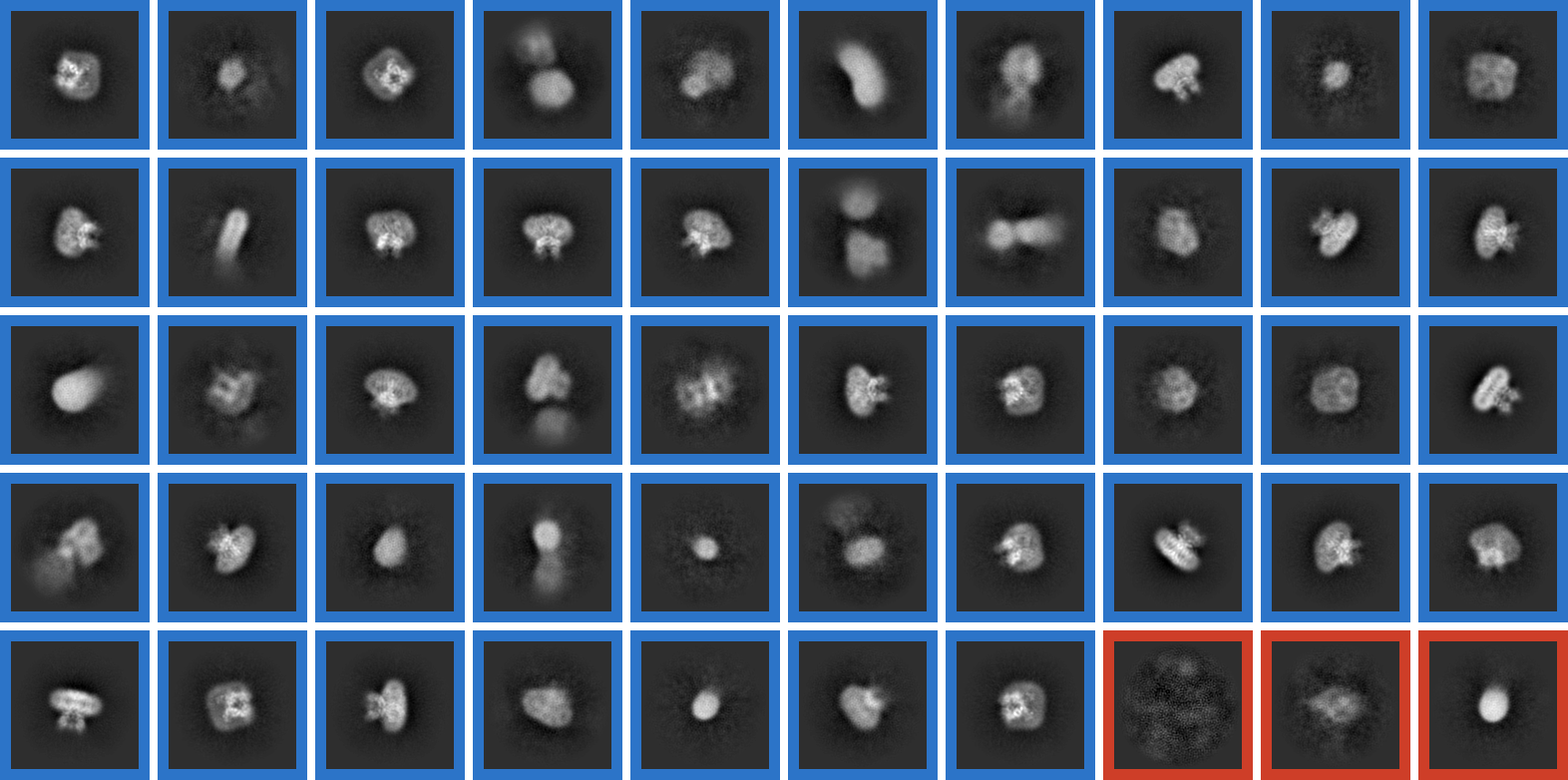}};
  \node[pan] (acf) at (13.50,-3.70) {\includegraphics[width=3.40cm]{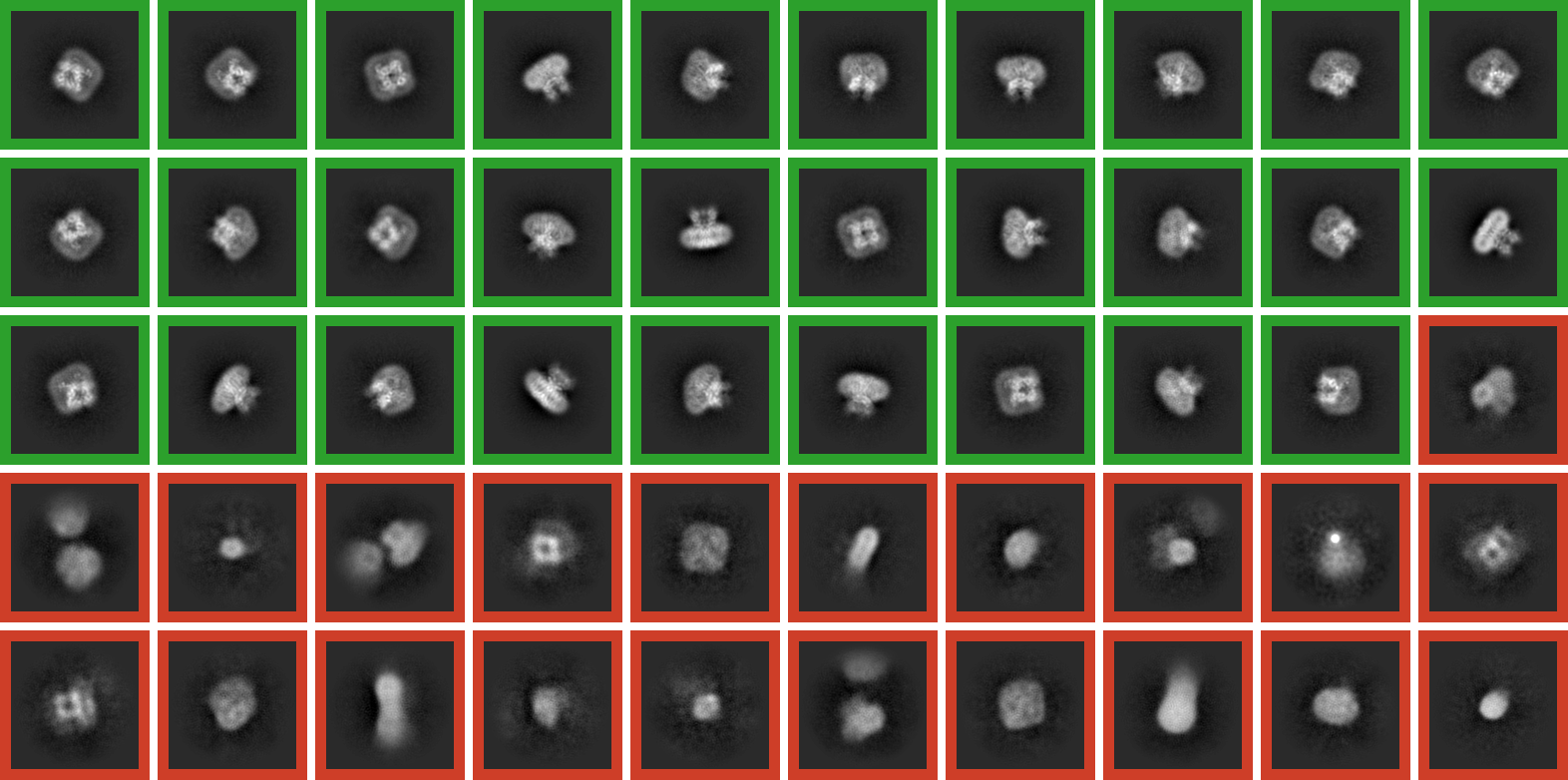}};

  \node[wlab] at (1.70,-4.62) {first classification, $K{=}50$};
  \node[wlab] at (6.20,-4.62) {cycle 1 classification};
  \node[wlab] at (10.70,-4.62) {cycle 2 classification};
  \node[wlab] at (15.20,-4.62) {final classification};

  \draw[ar] (ac0.east) -- (ac1.west);
  \draw[ar] (ac1.east) -- (ac2.west);
  \draw[ar] (ac2.east) -- (acf.west);
  \draw[ar, rounded corners=2pt] ([xshift=-5mm]ac0.north east) |- (8.45,-2.40)
        -| ([xshift=7.5mm]acf.north west);
  \node[font=\scriptsize, anchor=south] at (5.60,-2.28) {set aside, $s\le\ell$};

  \draw[ar, rounded corners=3pt] (a3.east) -- (14.45,0) -- (14.45,-1.72)
        -- (1.70,-1.72) -- (ac0.north);
  \draw[ar, rounded corners=3pt] (acf.north) -- (15.20,-2.20) -- (15.88,-2.20)
        -- (15.88,-1.30) -- (aft.south);
  \draw[ar, rounded corners=3pt] (aft.north) -- (15.88,1.15)
        -- node[above=0.4mm, font=\scriptsize, fill=white, inner sep=1pt, pos=0.3]
           {the next round picks with $\theta_{n+1}$}
        (1.18,1.15) -- (act.north);

  \node[car] (ap0) at (3.95,-3.70)  {\includegraphics[width=0.86cm]{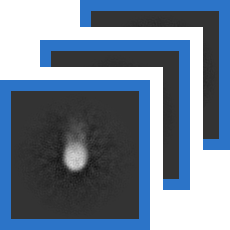}};
  \node[car] (ap1) at (8.45,-3.70)  {\includegraphics[width=0.86cm]{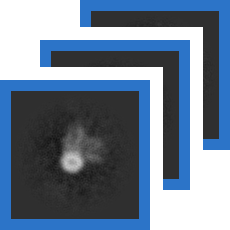}};
  \node[car] (ap2) at (12.95,-3.70) {\includegraphics[width=0.86cm]{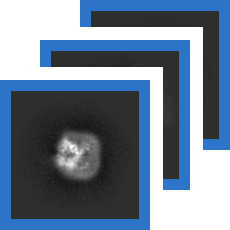}};
  \node[car] (aas) at (8.45,-2.40)  {\includegraphics[width=0.86cm]{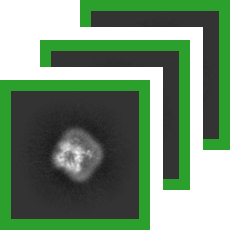}};
  \node[car] (atp) at (15.88,-1.30) {\includegraphics[width=0.86cm]{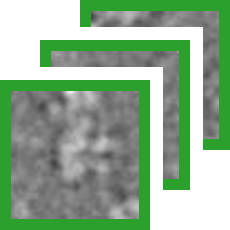}};

  \draw[sub] (-0.10,-1.90) rectangle (17.10,-5.15);
  \node[tag] at (0.05,-1.90) {2D class selection};
  \draw[blk] (-0.20,1.62) rectangle (17.20,-5.35);
  \node[tag] at (-0.05,1.62) {one round of the feedback loop, 300 annotated micrographs};

  \node[proc] (bct) at (0.10,-7.25)  {CryoTransformer};
  \node[pan]  (b1)  at (2.62,-7.25)  {\includegraphics[width=2.00cm]{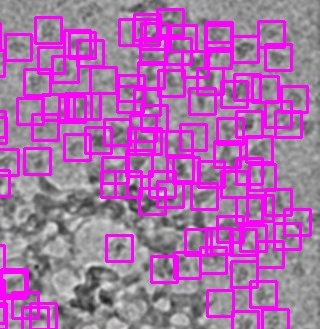}};
  \node[proc] (bmc) at (4.98,-7.25)  {MicrographCleaner};
  \node[pan]  (b2)  at (7.50,-7.25)  {\includegraphics[width=2.00cm]{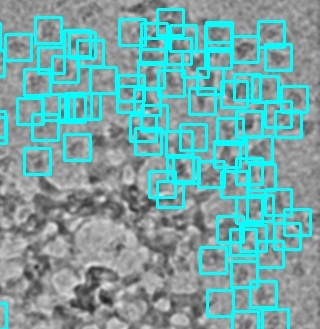}};
  \node[proc] (bex) at (9.86,-7.25)  {Extract};
  \node[pan]  (b3)  at (12.38,-7.25) {\includegraphics[width=1.70cm]{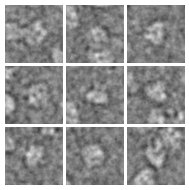}};

  \node[lab] at (3.62,-8.43) {picks};
  \node[lab] at (8.50,-8.43) {surviving picks};
  \node[lab] at (13.23,-8.43) {extracted particles};

  \draw[ar] (bct.east) -- (b1.west);
  \draw[ar] (b1.east) -- (bmc.west);
  \draw[ar] (bmc.east) -- (b2.west);
  \draw[ar] (b2.east) -- (bex.west);
  \draw[ar] (bex.east) -- (b3.west);

  \node[pan] (bc0) at (0.00,-10.95)  {\includegraphics[width=3.40cm]{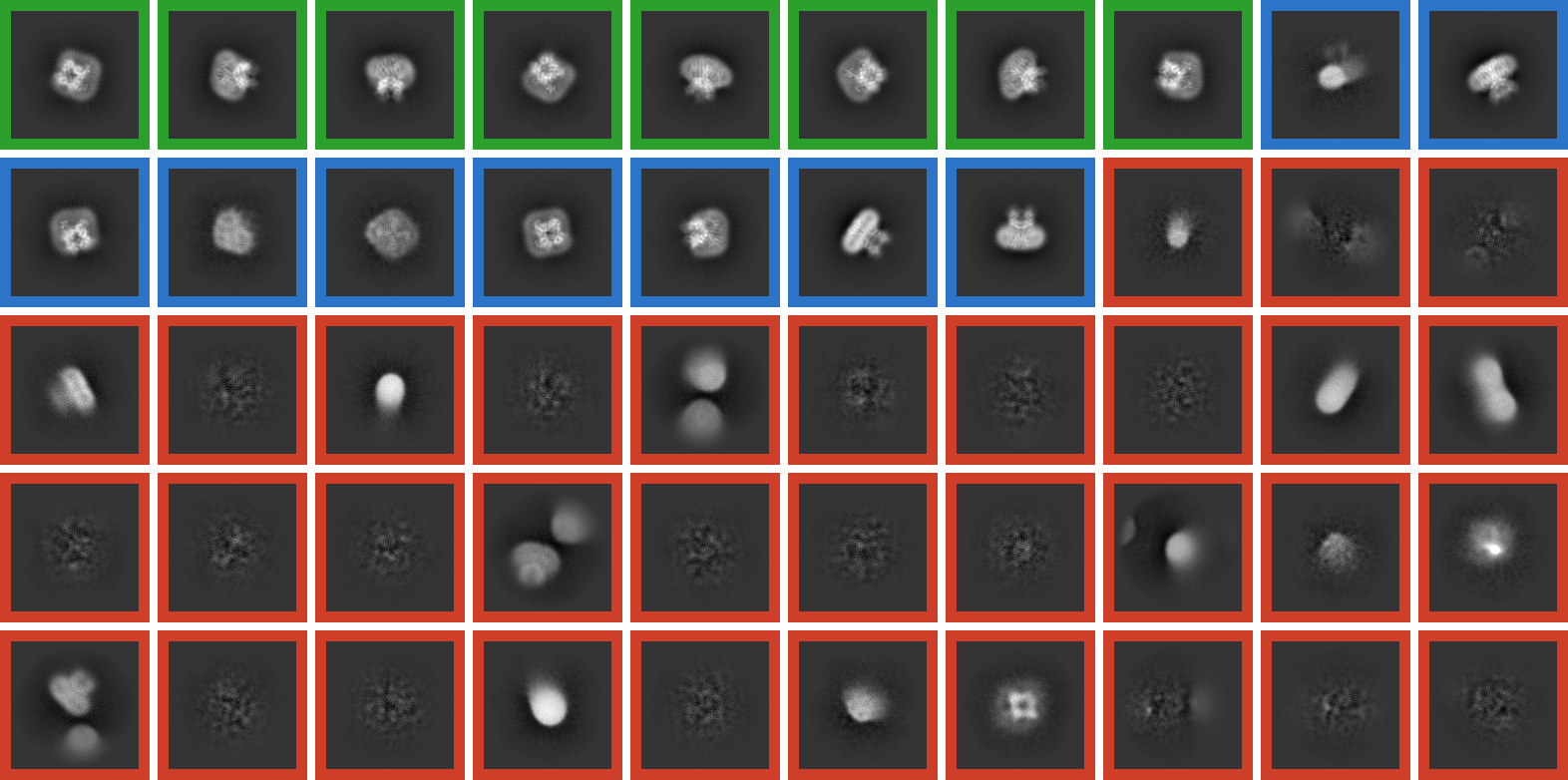}};
  \node[pan] (bc1) at (4.50,-10.95)  {\includegraphics[width=3.40cm]{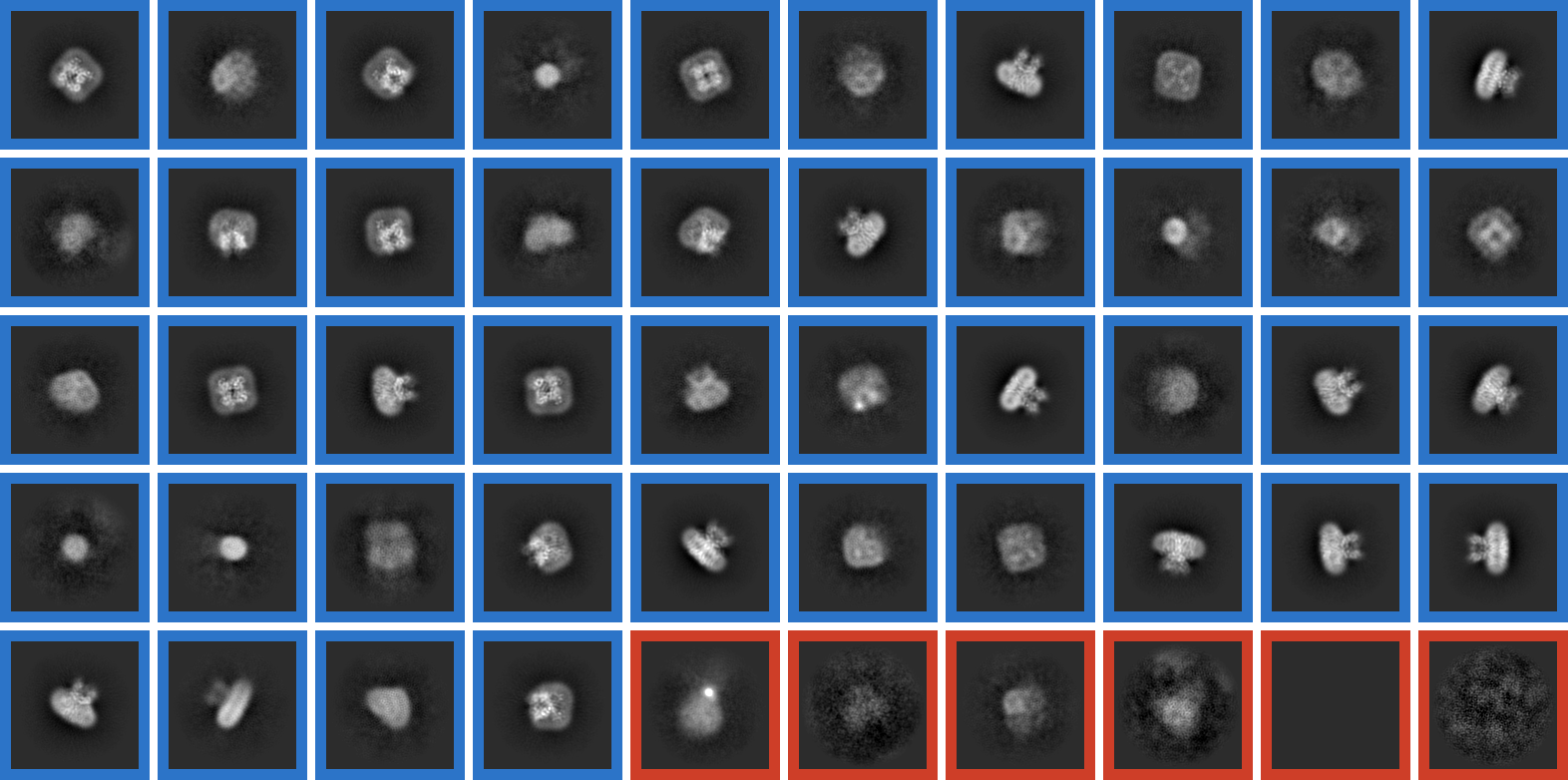}};
  \node[pan] (bc2) at (9.00,-10.95)  {\includegraphics[width=3.40cm]{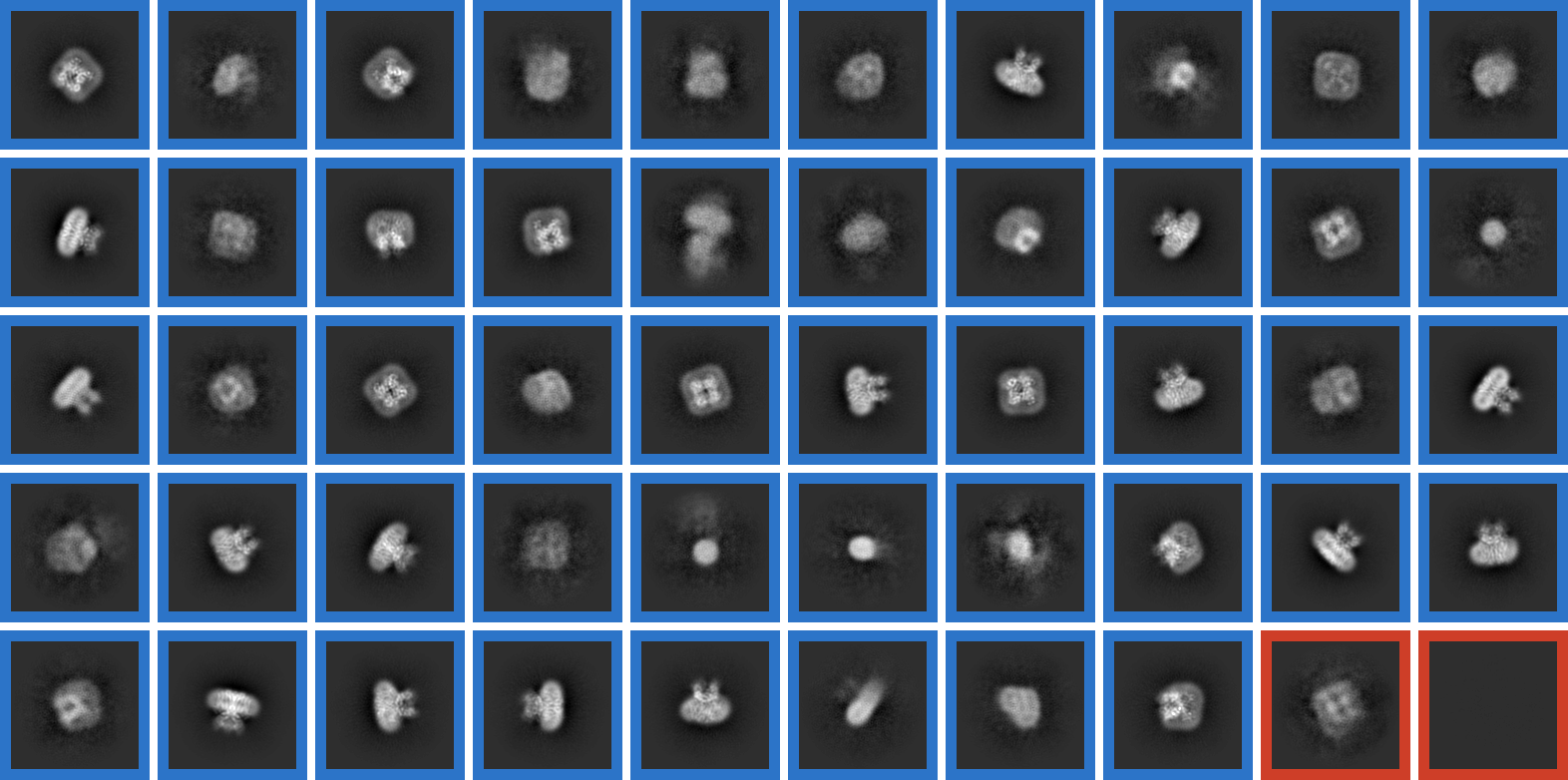}};
  \node[pan] (bcf) at (13.50,-10.95) {\includegraphics[width=3.40cm]{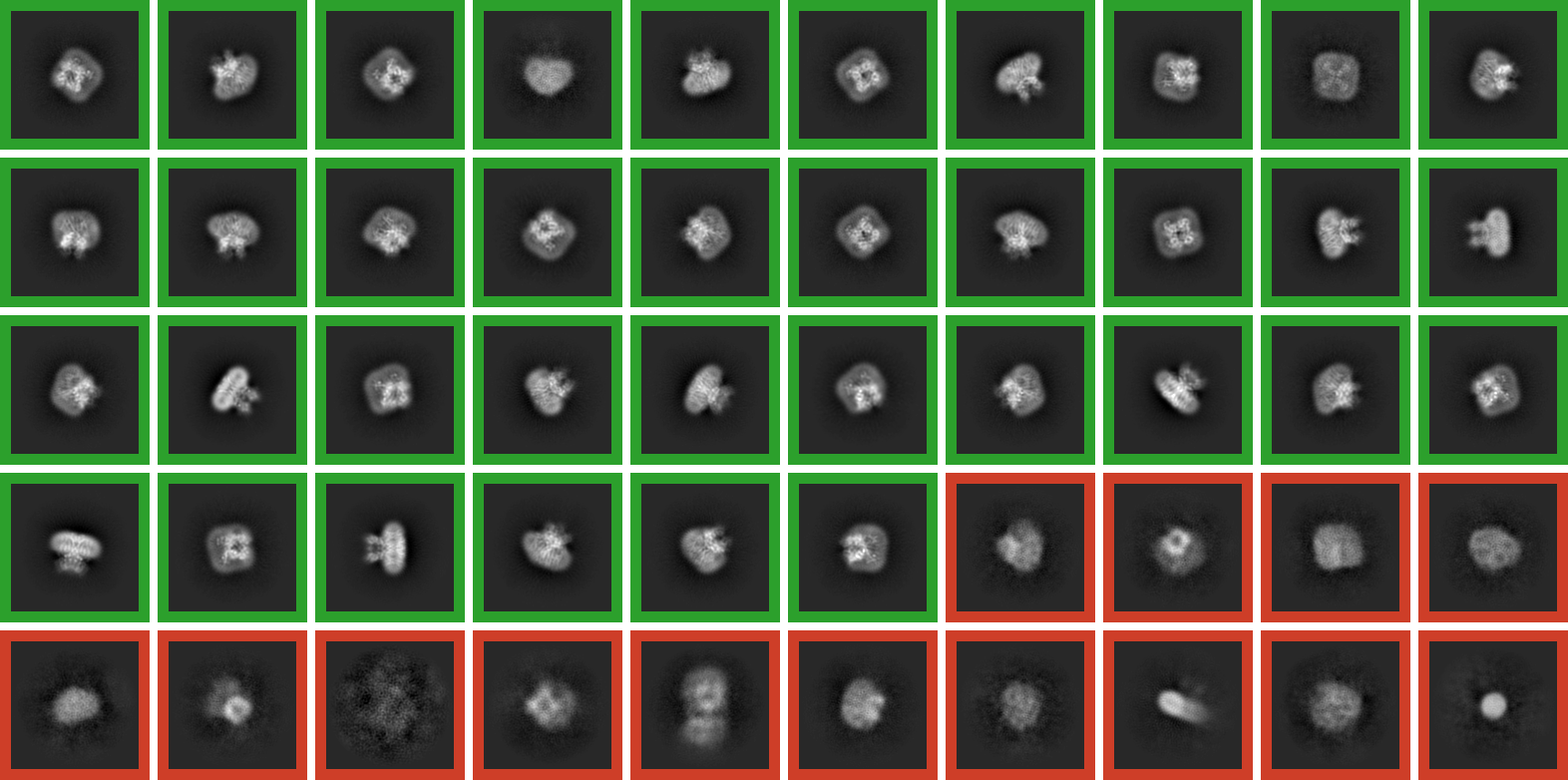}};

  \node[wlab] at (1.70,-11.87) {first classification, $K{=}50$};
  \node[wlab] at (6.20,-11.87) {cycle 1 classification};
  \node[wlab] at (10.70,-11.87) {cycle 2 classification};
  \node[wlab] at (15.20,-11.87) {final classification};

  \draw[ar] (bc0.east) -- (bc1.west);
  \draw[ar] (bc1.east) -- (bc2.west);
  \draw[ar] (bc2.east) -- (bcf.west);
  \draw[ar, rounded corners=2pt] ([xshift=-5mm]bc0.north east) |- (8.45,-9.65)
        -| ([xshift=7.5mm]bcf.north west);
  \node[font=\scriptsize, anchor=south] at (5.60,-9.53) {set aside, $s\le\ell$};

  \draw[ar, rounded corners=3pt] (b3.east) -- (14.45,-7.25) -- (14.45,-8.97)
        -- (1.70,-8.97) -- (bc0.north);

  \node[car] (bp0) at (3.95,-10.95)  {\includegraphics[width=0.86cm]{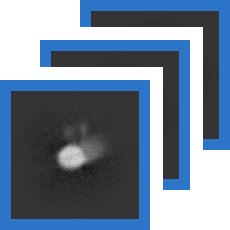}};
  \node[car] (bp1) at (8.45,-10.95)  {\includegraphics[width=0.86cm]{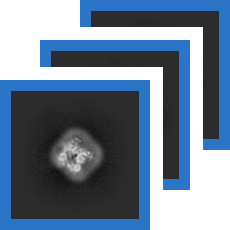}};
  \node[car] (bp2) at (12.95,-10.95) {\includegraphics[width=0.86cm]{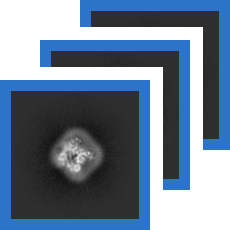}};
  \node[car] (bas) at (8.45,-9.65)   {\includegraphics[width=0.86cm]{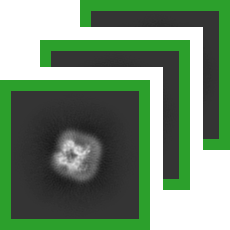}};

  \draw[sub] (-0.10,-9.15) rectangle (17.10,-12.40);
  \node[tag] at (0.05,-9.15) {2D class selection};

  \node[pan] (c1) at (0.35,-14.00)  {\includegraphics[width=2.80cm]{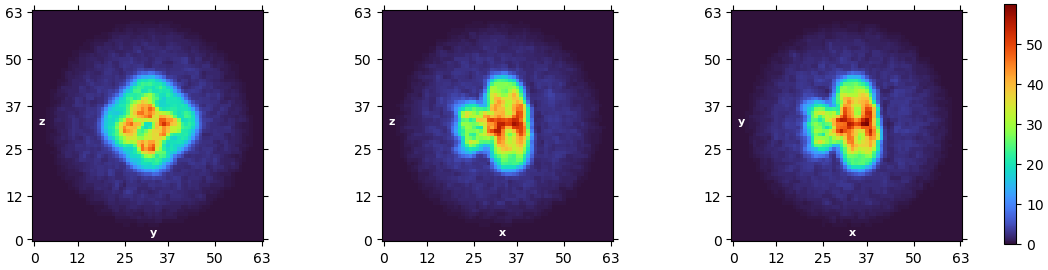}};
  \node[pan] (c2) at (3.70,-14.00)  {\includegraphics[width=3.40cm]{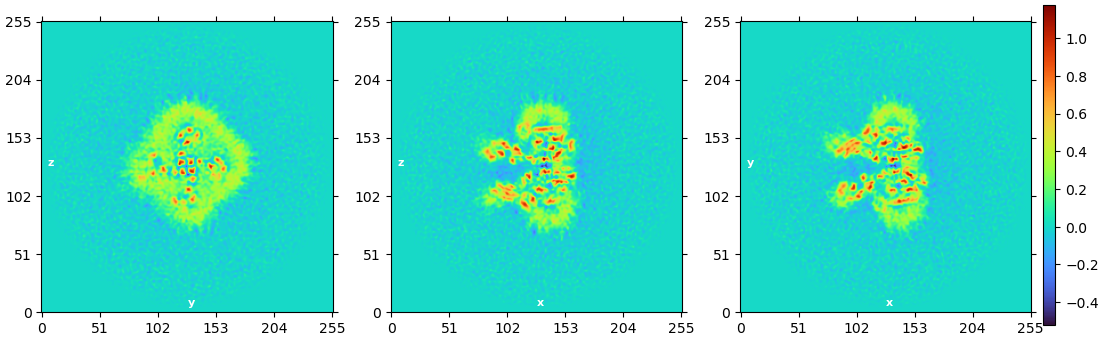}};
  \node[pan] (c3) at (7.65,-14.00)  {\includegraphics[width=2.60cm]{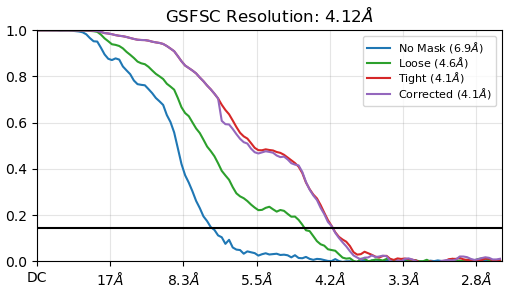}};
  \node[pan] (c4) at (10.80,-14.00) {\includegraphics[width=3.60cm]{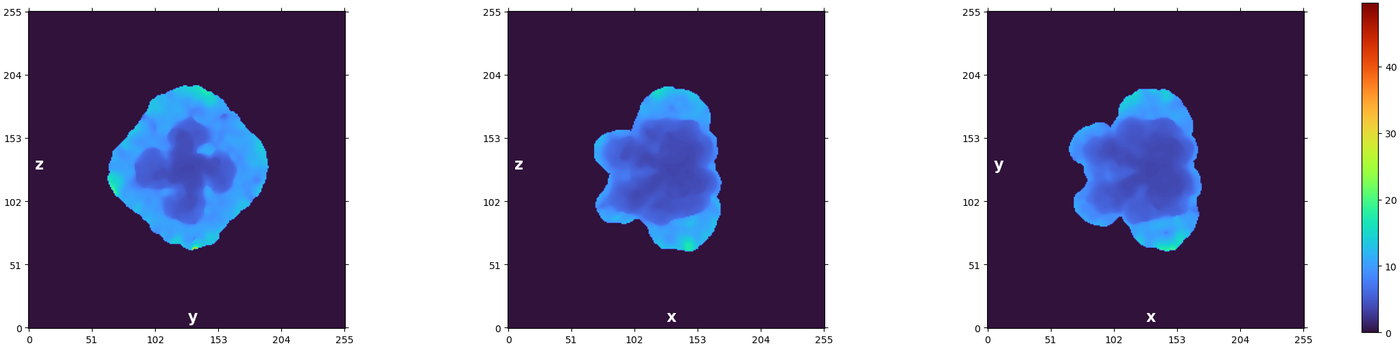}};
  \node[pan] (c5) at (14.95,-14.00) {\includegraphics[width=2.20cm]{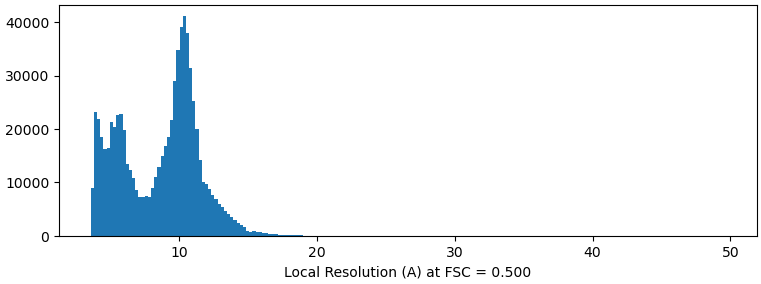}};

  \node[wlab] at (1.75,-14.87) {Ab-initio Reconstruction\\three seeds};
  \node[wlab] at (5.40,-14.87) {Homogeneous Refinement\\three seeds};
  \node[wlab] at (8.95,-14.87) {GSFSC, best of the three};
  \node[wlab] at (13.55,-14.87) {Local Resolution Estimation};

  \draw[ar] (c1.east) -- (c2.west);
  \draw[ar] (c2.east) -- (c3.west);
  \draw[ar] (c3.east) -- (c4.west);
  \draw[ar, rounded corners=3pt] (bcf.east) -- (16.95,-10.95) -- (16.95,-12.90)
        -- (1.75,-12.90) -- (c1.north);
  \node[car] (btp) at (5.40,-12.90)  {\includegraphics[width=0.86cm]{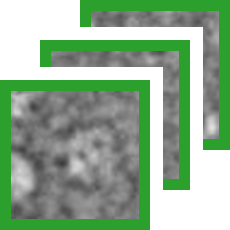}};

  \draw[blk] (-0.20,-6.05) rectangle (17.20,-15.70);
  \node[tag] at (-0.05,-6.05) {the full micrograph set};

  \draw[ar] (1.18,-5.35) -- node[right, font=\scriptsize, pos=0.2]
        {the checkpoint the loop delivers picks the full micrograph set} (bct.north);

  \node[font=\scriptsize, anchor=north] at (8.50,-16.05)
       {\textcolor{pcaside}{\rule{6pt}{6pt}}~kept: set aside at the first classification,
        or delivered by the final one \quad
        \textcolor{pcpool}{\rule{6pt}{6pt}}~re-classified in the next cycle \quad
        \textcolor{pcdrop}{\rule{6pt}{6pt}}~discarded};
  \end{tikzpicture}
  \caption{\textbf{The pipeline as it runs.} One round of the feedback loop on the 300
  annotated micrographs (upper block) and the full micrograph set picked with the
  checkpoint the loop delivers (lower block), every panel as CryoSPARC renders it for that
  job. \Cref{sec:supp:protocol} describes each stage.}
  \label{fig:protocol}
\end{figure*}

\Cref{fig:protocol} draws the pipeline of the main paper on one of the datasets. The upper block is one round of the feedback loop on the 300 annotated micrographs. The lower block picks the full micrograph set with the checkpoint the loop delivers and carries that set through to a local-resolution estimate. Both blocks begin with CryoTransformer picking the candidates, MicrographCleaner removing the picks that fall on contamination, and CryoSPARC extracting what survives.

The dotted box of each block is the 2D class selection stage, drawn one classification at a time. Each panel is the 50 class averages of one classification, framed by what the selection does to each class and grouped by that fate. Every arrow of the stage carries three of the class averages that travel along it. The blue classes of one panel are the particles the next panel classifies, so the red block shrinks from cycle to cycle as the pool runs out of classes to discard. The green classes of the first classification skip the cycles and rejoin at the final classification.

The round of the loop runs no reconstruction. It ends at the classes the final selection keeps, whose particles on 50 of the micrographs become the pseudo-labels that fine-tune the picker. On the full set, the particles of those classes go on to ab-initio reconstruction and homogeneous refinement. Both are particle coordinates rather than class averages, so the two arrows leaving the stage carry raw particles cropped from the extraction render. These stages run three times with different random seeds, and local resolution is estimated on the best of the three refined maps by GSFSC resolution. The resolution burnt into the gold-standard FSC panel is the number the main paper reports for this entry.

\section{Details of the picker and its fine-tuning}
\label{sec:supp:training}

\begin{table}[t]
  \centering
  \small
  \caption{\textbf{Settings of the two training stages.} Head repair fits the
  classification head alone on the features of the frozen detector, and fine-tuning
  trains every weight on the teacher set of one round. Each of the three weights scales one
  term of the loss, and the no-object weight scales the classification term on the queries
  labeled no-object.}
  \label{tab:supp_hparams}
  \setlength{\tabcolsep}{1pt}
  \begin{tabular*}{\linewidth}{@{\extracolsep{\fill}}lll@{}}
    \toprule
     & head repair & fine-tuning \\
    \midrule
    trained weights  & head only & all \\
    training set     & 22 CryoPPP entries & 40 micrographs \\
    loss                     & two-class softmax CE & Hungarian set loss \\
    \quad class weight     & 1 & 1 \\
    \quad $L_1$ box weight & -- & 5 \\
    \quad GIoU weight      & -- & 2 \\
    \quad no-object weight      & 0.1 & 0.1 \\
    auxiliary decoder losses & -- & 5 layers \\
    optimizer        & Adam & AdamW \\
    learning rate    & $2 \times 10^{-3}$ & $10^{-4}$ \\
    backbone lr      & -- & $10^{-5}$ \\
    lr decay         & -- & $\times 0.1$ at epoch 24 \\
    weight decay     & $10^{-4}$ & $10^{-4}$ \\
    gradient clip    & -- & 0.1 \\
    batch size       & 32{,}768 queries & 8 micrographs \\
    epochs           & 15 & 50 \\
    \bottomrule
  \end{tabular*}
\end{table}

\noindent\textbf{Operating point.} CryoTransformer emits a fixed set of 600 scored candidate queries per micrograph. We keep the operating point of the original implementation. Each micrograph keeps the top 75\% of candidates by score, and duplicates are removed by non-maximum suppression at an overlap threshold of 0.7.

\noindent\textbf{Base picker checkpoint.} The classification head of CryoTransformer is a single linear layer over the 256-dimensional output of the last decoder layer. The head repair of the main paper discards the released weights of that layer and retrains it on the features the frozen detector produces for the 22 CryoPPP entries \cite{cryoppp2023} of the picker's training set. The features are standardized dimension by dimension for training, which puts the 256 dimensions on a common scale, and the standardization is folded back into the retrained weights afterwards, so that inference runs the detector unchanged. The layer is then written back into the two-class head with the weight and the bias of the no-object class set to zero. We hold out no split here, so the deployed head is the one fit on all micrographs of the 22 entries.

\noindent\textbf{Fine-tuning.} Each round of the feedback loop restarts from the repaired base picker checkpoint and trains every weight of the picker on the training micrographs of that round's teacher set. The detection loss is the Hungarian set loss of CryoTransformer \cite{cryotransformer2024}, taken unchanged. The number of epochs and the epoch of the decay follow the fine-tuning stage of UPicker \cite{upicker2024}, and the remaining values of \cref{tab:supp_hparams}, the loss weights included, are the defaults of CryoTransformer. Each round delivers the weights of its last epoch, and the 10 validation micrographs monitor the loss only.

\section{Details of contamination masking}
\label{sec:supp:maskpostproc}

\begin{figure*}[t]
  \centering
  \includegraphics[width=0.8\linewidth]{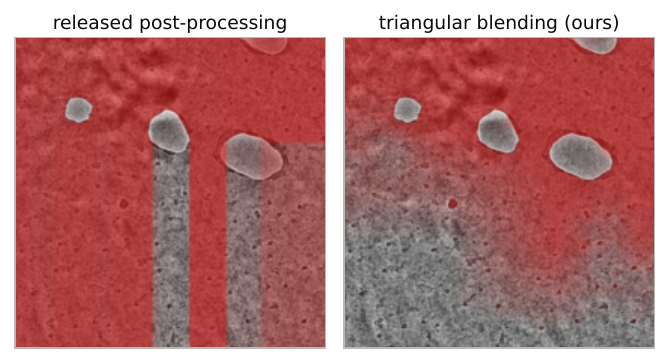}
  \caption{\textbf{The two post-processings on a real micrograph.} The contamination
  mask of one EMPIAR-10532 micrograph under the released post-processing of
  MicrographCleaner (left) and under our triangular blending (right), from the same
  network on the same input, drawn in red over the denoised micrograph. In the released
  mask the unmasked area survives only as vertical stripes with straight edges that run
  to the bottom of the frame. The triangular mask draws the same field with a smooth
  boundary.}
  \label{fig:mask_real}
\end{figure*}

\begin{figure}[t]
  \centering
  \includegraphics[width=\linewidth]{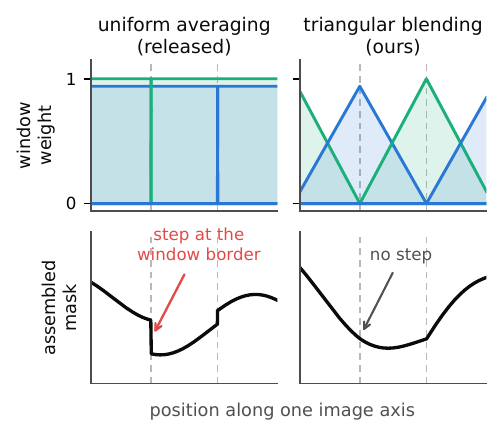}
  \caption{\textbf{Uniform averaging against triangular blending.} Four windows at 50\%
  overlap predict the same field with window-specific offsets. Top, the weight each window
  contributes along one image axis. Bottom, the assembled mask. A uniform window
  contributes at full weight up to its own border, so the assembled mask steps wherever
  adjacent windows disagree, and the released implementation repairs these steps
  afterwards. A triangular window hands over smoothly, so no step forms and no repair is
  needed.}
  \label{fig:blend_weights}
\end{figure}

MicrographCleaner \cite{micrographcleaner2020} predicts contamination on fixed-size windows, so the per-window predictions must be assembled into one mask per micrograph. \Cref{fig:mask_real} compares the released assembly with ours on one micrograph.

The released assembly averages the overlapping windows with uniform weights. A window contributes at full weight up to its own border, so the assembled mask carries a step wherever adjacent windows disagree (\cref{fig:blend_weights}, left). The routine \texttt{fixJumpInBorders} repairs these steps afterwards. When the mask falls sharply across a window border along a large fraction of the border, the routine overwrites everything between that border and the image edge with the values just before the border. The mask also falls this sharply where the micrograph itself changes intensity steeply. The routine cannot tell the two apart. When it misreads such a fall, the overwrite floods a rectangular region of the mask. The released assembly also averages the mask over eight rotations of the input.

We replace this assembly with triangular blending. Windows overlap by half their size, and the predictions are averaged with weights that fall linearly from one at the window center to zero at the border. Adjacent windows therefore hand over smoothly and no step forms, even where they disagree (\cref{fig:blend_weights}, right). This removes the need for a seam correction. We also drop the rotation averaging and predict each micrograph in a single orientation.

\section{Details of 2D class selection}
\label{sec:supp:cryosift}

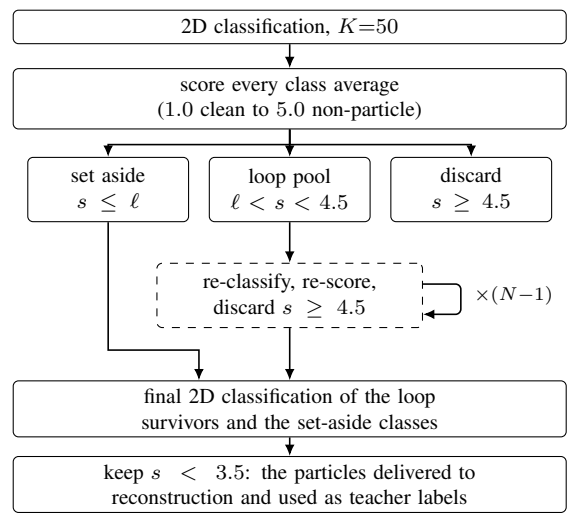
\begin{figure}[t]
  \centering
  \begin{tikzpicture}[
    font=\footnotesize, x=1cm, y=1cm,
    >={Latex[length=1.5mm,width=1.3mm]},
    stage/.style={draw, rounded corners=2pt, align=center, inner sep=3pt,
                  text width=7.1cm},
    fate/.style={draw, rounded corners=2pt, align=center, inner sep=2.5pt,
                 text width=1.95cm, minimum height=8.5mm},
    lp/.style={draw, dashed, rounded corners=2pt, align=center, inner sep=3pt,
               text width=3.3cm, minimum height=8.5mm},
    ar/.style={->, semithick},
  ]
  \node[stage] (c2d)   at (0,0)     {2D classification, $K{=}50$};
  \node[stage] (score) at (0,-0.95) {score every class average\\($1.0$ clean to $5.0$ non-particle)};
  \node[fate] (aside) at (-2.4,-2.15) {set aside\\$s\le\ell$};
  \node[fate] (pool)  at (0,-2.15)    {loop pool\\$\ell<s<4.5$};
  \node[fate] (drop)  at (2.4,-2.15)  {discard\\$s\ge4.5$};
  \node[lp]    (loop)  at (0,-3.55) {re-classify, re-score,\\discard $s\ge4.5$};
  \node[stage] (final) at (0,-5.05) {final 2D classification of the loop survivors and the set-aside classes};
  \node[stage] (keep)  at (0,-6.05) {keep $s<3.5$: the particles delivered to reconstruction and used as teacher labels};
  \draw[ar] (c2d) -- (score);
  \draw[ar] (score.south) -- ++(0,-0.2) -| (aside.north);
  \draw[ar] (score.south) -- ++(0,-0.2) -| (drop.north);
  \draw[ar] (score) -- (pool);
  \draw[ar] (pool) -- (loop);
  \draw[ar, rounded corners=2pt] ([yshift=1.5mm]loop.east) -- ++(0.5,0)
        -- ++(0,-0.4) -- ([yshift=-2.5mm]loop.east);
  \node[anchor=west, font=\scriptsize] at ([xshift=5.5mm]loop.east) {$\times(N{-}1)$};
  \draw[ar] (loop) -- (final);
  \draw[ar] (aside.south) |- ([yshift=0.4cm,xshift=-1.2cm]final.north) -- ++(0,-0.4);
  \draw[ar] (final) -- (keep);
  \end{tikzpicture}
  \caption{\textbf{The iterative workflow of CryoSift.} The best classes are set aside at
  the threshold $\ell$, the worst are discarded at $4.5$, and the rest are re-classified for
  $N$ cycles under the same rule. A final classification over the survivors and the
  set-aside classes delivers the classes scoring better than $3.5$.}
  \label{fig:cryosift_flow}
\end{figure}

\Cref{fig:cryosift_flow} draws the 2D class selection stage of the main paper as we run it. Every class average of the first classification carries a score $s$, from 1.0 for clean particle classes to 5.0 for non-particle classes, and the classes split three ways on it. Classes scoring 2.5 or better are candidates for setting aside, and the best 70\% of them, counted in classes, are held out of the loop so that they cannot attract particles in the cycles that follow. The split threshold $\ell$ is the score of the class at that mark. Classes scoring 4.5 or worse are discarded permanently, and the rest enter the loop, where each cycle re-classifies the survivors and again discards whatever reaches 4.5. A cycle that finds nothing at 4.5 or worse ends the loop early. The set-aside classes then rejoin the loop survivors for a final classification, and the classes scoring better than 3.5 there are what the stage delivers. CryoSift's box-size rule \cite{cryosift2025} sets the number of cycles $N$. The extraction boxes of all four datasets fall in the 200 to 300 pixel bracket, giving $N=3$.

The 2D classification runs at $K=50$, roughly two thousand particles per class and 20 to 40 times coarser than ISAC \cite{isac2012}, so non-particles that fall in the same class as genuine particles survive the selection. Keeping or discarding whole classes cannot separate the two, which limits the precision of the teacher and therefore the precision the picker can learn.

\section{The 2D selection failure on EMPIAR-10345}
\label{sec:supp:selection}

\begin{figure}[t]
  \centering
  \definecolor{fcaside}{RGB}{44,160,44}
  \definecolor{fcpool}{RGB}{44,116,200}
  \definecolor{fcdrop}{RGB}{206,62,40}
  \setlength{\tabcolsep}{0pt}
  \begin{tabular}{c}
    \footnotesize EMPIAR-10081 \\
    \includegraphics[width=0.98\linewidth]{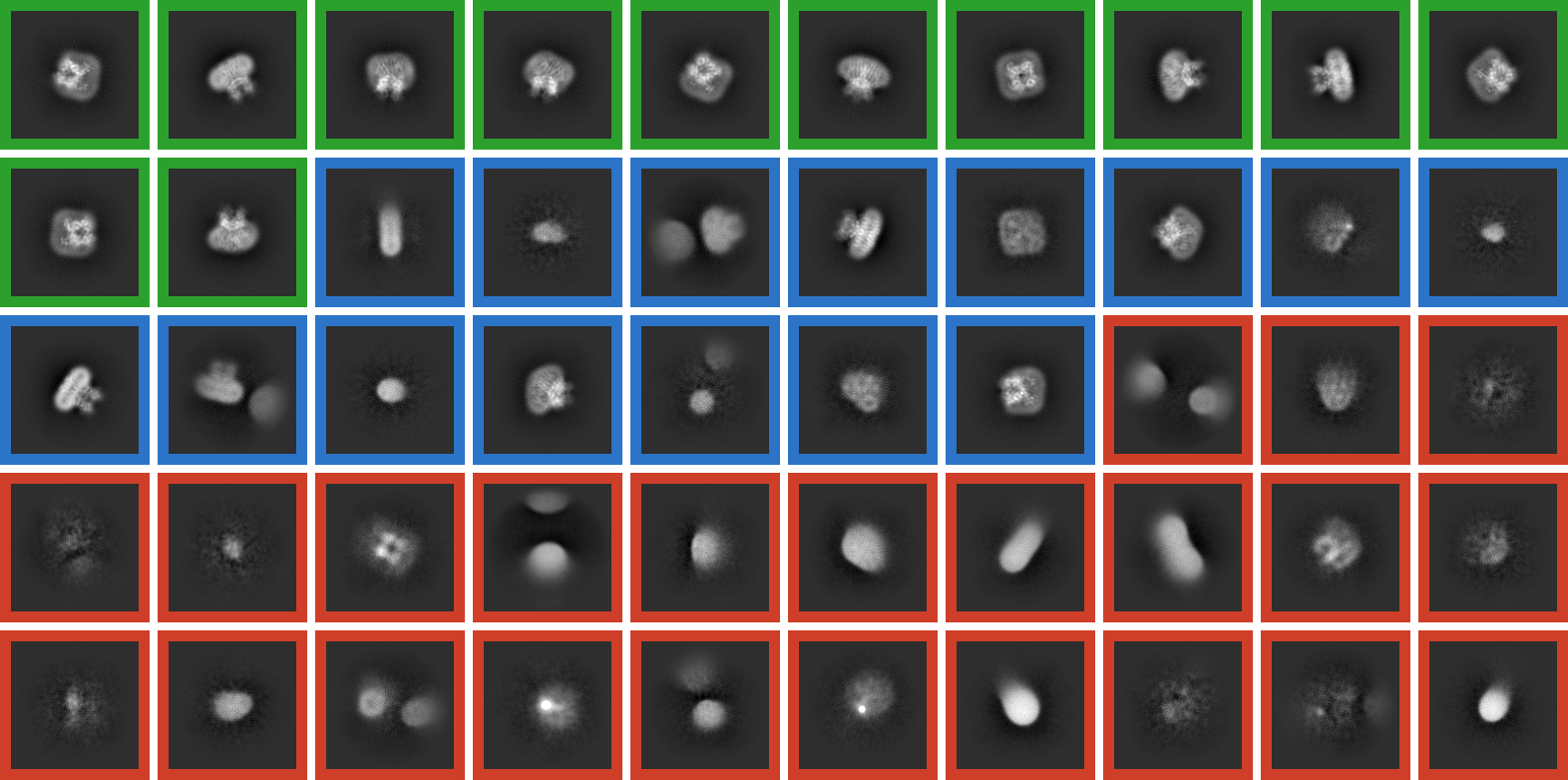} \\[4pt]
    \footnotesize EMPIAR-10345 \\
    \includegraphics[width=0.98\linewidth]{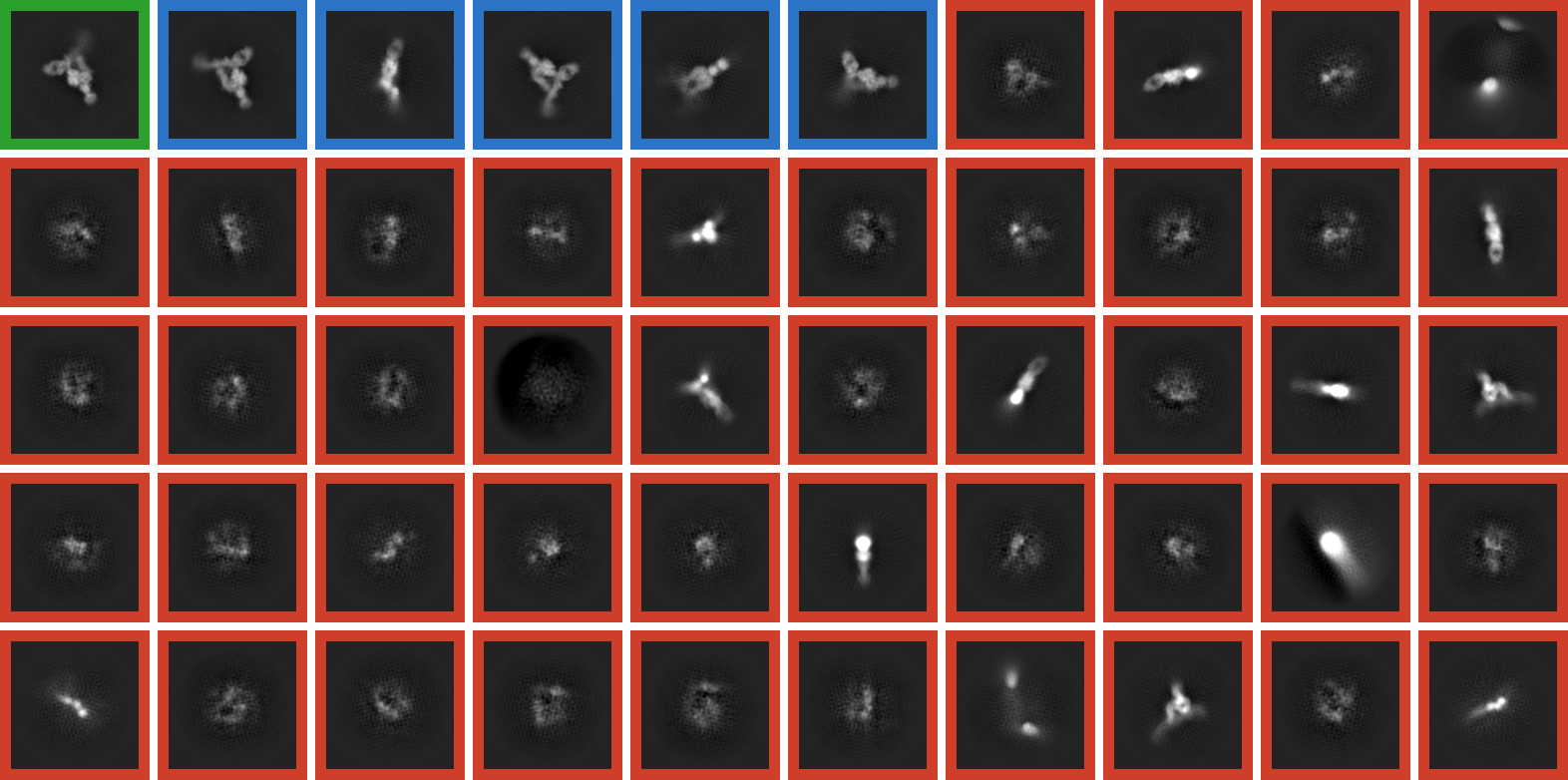} \\
  \end{tabular}
  \\[3pt]
  {\footnotesize
   \textcolor{fcaside}{\rule{7pt}{7pt}}~set aside, $s\le\ell$ \quad
   \textcolor{fcdrop}{\rule{7pt}{7pt}}~discarded, $s\ge4.5$ \\
   \textcolor{fcpool}{\rule{7pt}{7pt}}~re-classified in the next cycle, $\ell<s<4.5$}
  \caption{\textbf{What the first cycle of 2D class selection does.} The 50 class averages
  of the first classification of each full set, colored by what the first cycle of
  \cref{fig:cryosift_flow} does to each class and grouped by that fate. The cycle discards
  23 of the 50 classes on EMPIAR-10081 and 44 on EMPIAR-10345, and a discarded class is
  gone for good.}
  \label{fig:first_cycle}
\end{figure}

\Cref{fig:first_cycle} shows where EMPIAR-10345 loses its particles. Under contamination masking followed by 2D class selection, which removes 94.1\% of the candidates of that entry, the first cycle discards 44 of the 50 classes, against 23 on EMPIAR-10081. The later cycles re-classify only what survives, so this first cut bounds the entry (\cref{sec:supp:cryosift}).

\section{Diagnostics of main results and ablation}
\label{sec:supp:diagnostics}

\begin{figure*}[!t]
  \centering
  \setlength{\tabcolsep}{1.5pt}
  \renewcommand{\arraystretch}{0.9}
  \begin{tabular}{lcccc}
   & \footnotesize EMPIAR-10081 & \footnotesize EMPIAR-10093 & \footnotesize EMPIAR-10345 & \footnotesize EMPIAR-10532 \\
    \parbox[b][0.120\linewidth][c]{2.0cm}{\footnotesize\raggedright crYOLO\\{\scriptsize full set}} &
    \includegraphics[width=0.207\linewidth]{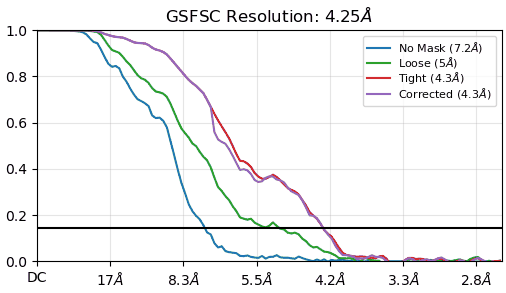} &
    \includegraphics[width=0.207\linewidth]{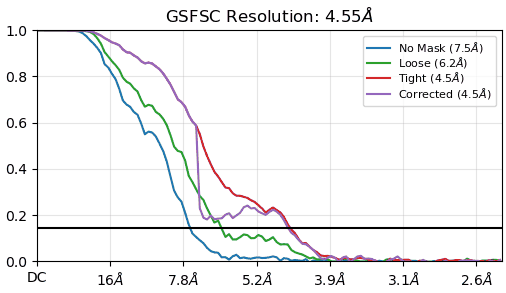} &
    \includegraphics[width=0.207\linewidth]{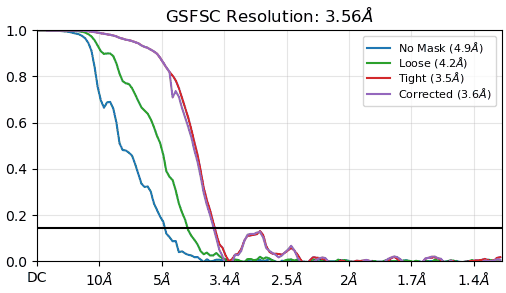} &
    \includegraphics[width=0.207\linewidth]{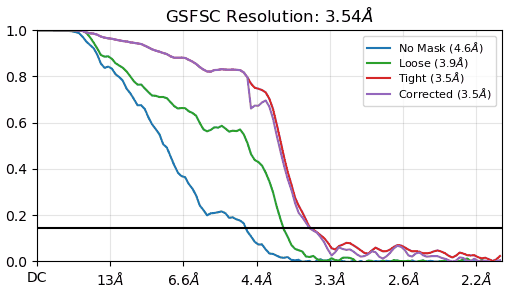} \\
    \parbox[b][0.120\linewidth][c]{2.0cm}{\footnotesize\raggedright Topaz\\{\scriptsize full set}} &
    \includegraphics[width=0.207\linewidth]{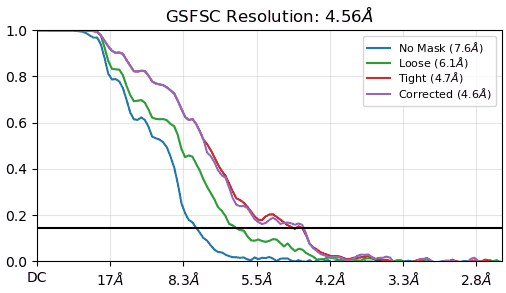} &
    \includegraphics[width=0.207\linewidth]{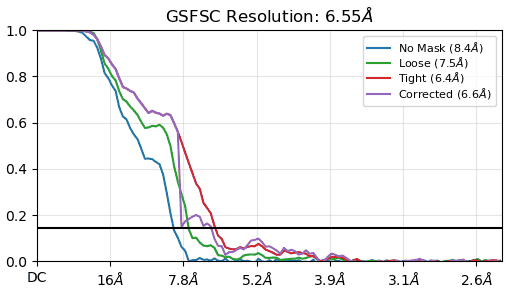} &
    \includegraphics[width=0.207\linewidth]{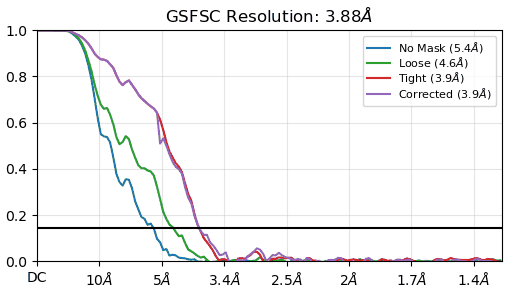} &
    \includegraphics[width=0.207\linewidth]{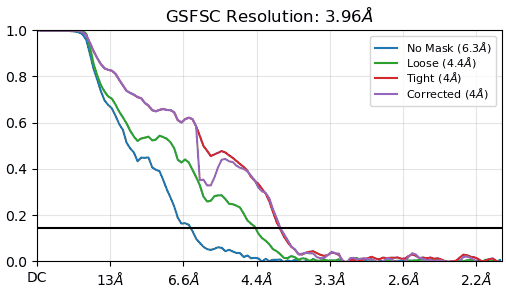} \\
    \parbox[b][0.120\linewidth][c]{2.0cm}{\footnotesize\raggedright CryoSegNet\\{\scriptsize full set}} &
    \includegraphics[width=0.207\linewidth]{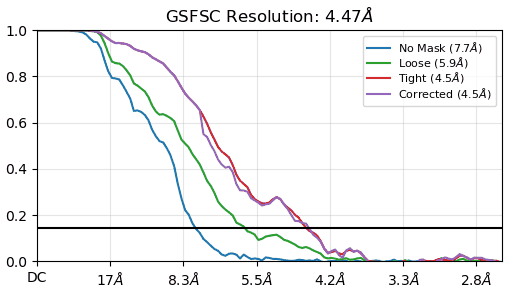} &
    \includegraphics[width=0.207\linewidth]{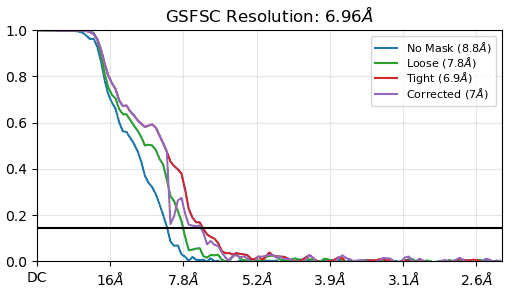} &
    \includegraphics[width=0.207\linewidth]{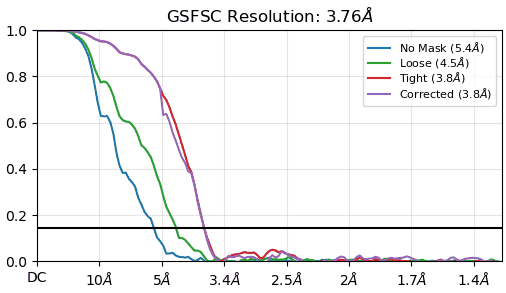} &
    \includegraphics[width=0.207\linewidth]{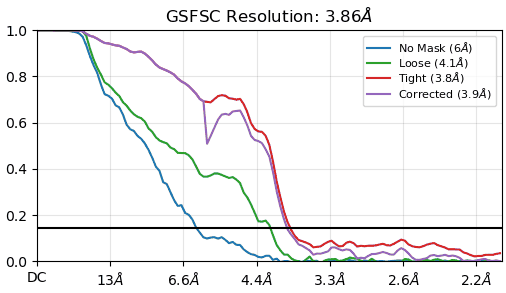} \\
    \parbox[b][0.120\linewidth][c]{2.0cm}{\footnotesize\raggedright CryoTransformer\\(baseline)\\{\scriptsize full set}} &
    \includegraphics[width=0.207\linewidth]{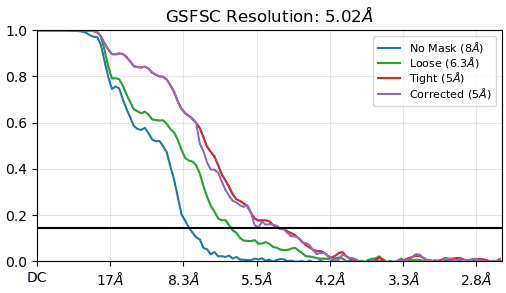} &
    \includegraphics[width=0.207\linewidth]{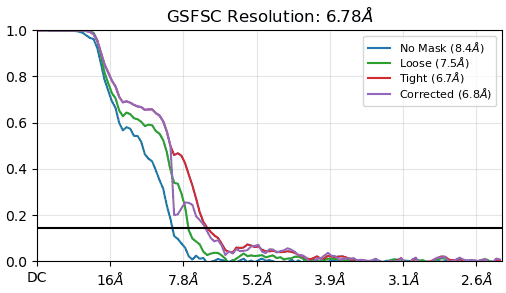} &
    \includegraphics[width=0.207\linewidth]{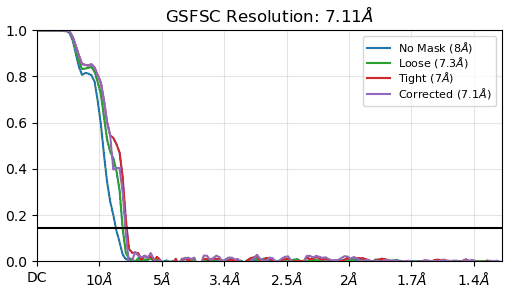} &
    \includegraphics[width=0.207\linewidth]{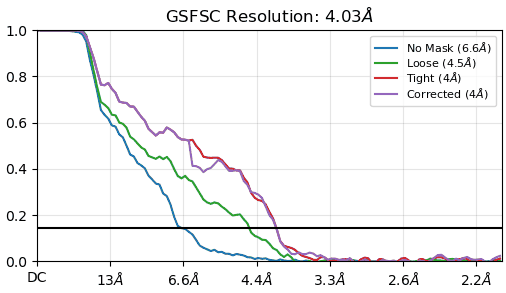} \\
    \parbox[b][0.120\linewidth][c]{2.0cm}{\footnotesize\raggedright $+$mask\\{\scriptsize full set}} &
    \includegraphics[width=0.207\linewidth]{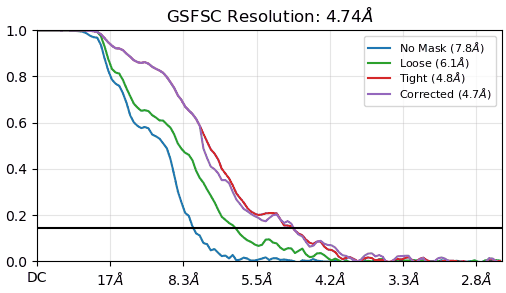} &
    \includegraphics[width=0.207\linewidth]{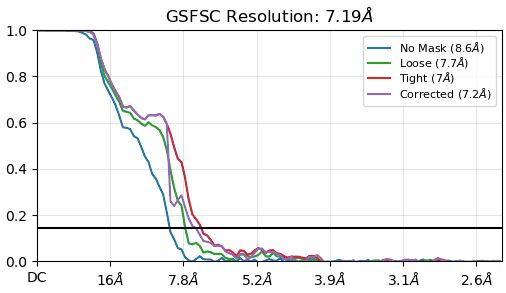} &
    \includegraphics[width=0.207\linewidth]{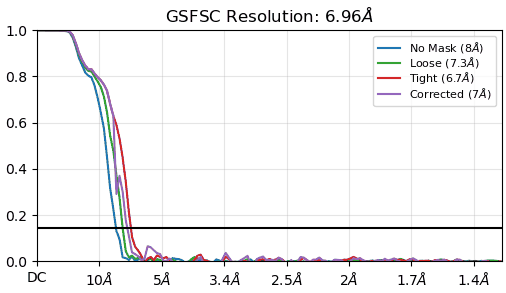} &
    \includegraphics[width=0.207\linewidth]{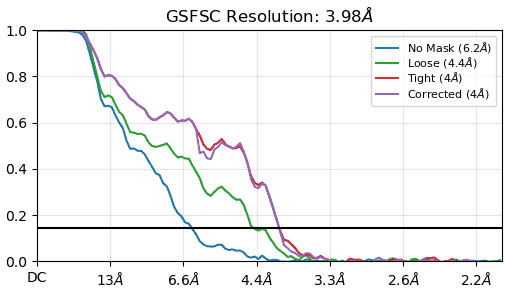} \\
    \parbox[b][0.120\linewidth][c]{2.0cm}{\footnotesize\raggedright $+$select\\{\scriptsize full set}} &
    \includegraphics[width=0.207\linewidth]{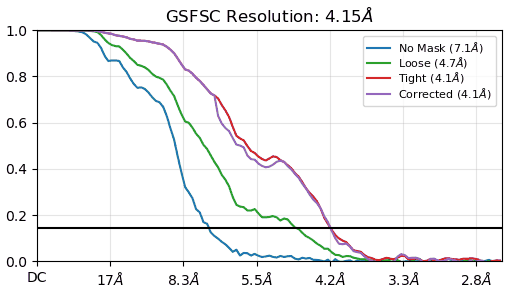} &
    \includegraphics[width=0.207\linewidth]{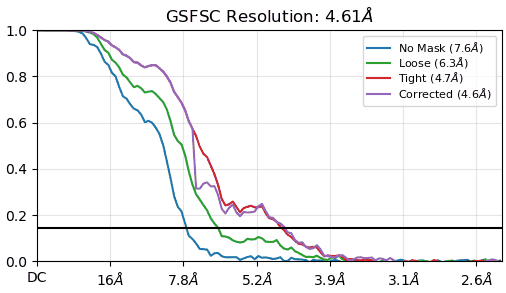} &
    \includegraphics[width=0.207\linewidth]{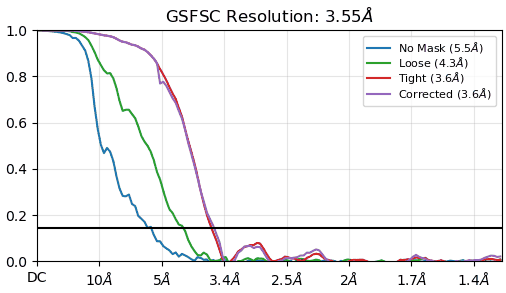} &
    \includegraphics[width=0.207\linewidth]{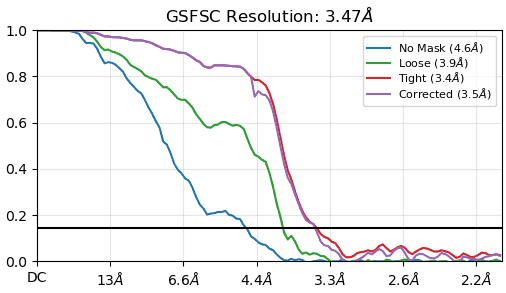} \\
    \parbox[b][0.120\linewidth][c]{2.0cm}{\footnotesize\raggedright $+$both\\{\scriptsize full set}} &
    \includegraphics[width=0.207\linewidth]{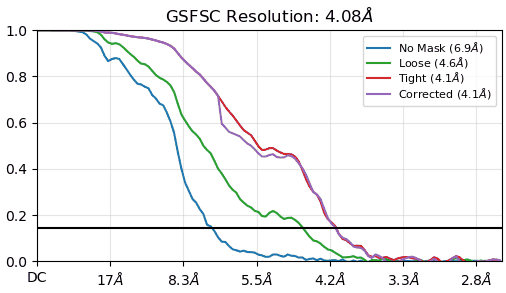} &
    \includegraphics[width=0.207\linewidth]{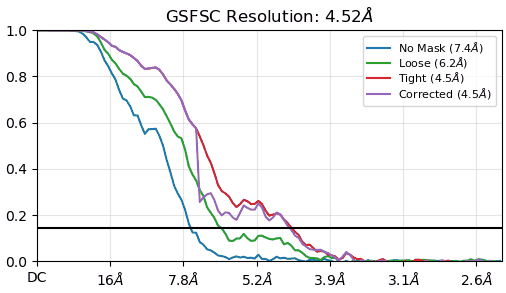} &
    \includegraphics[width=0.207\linewidth]{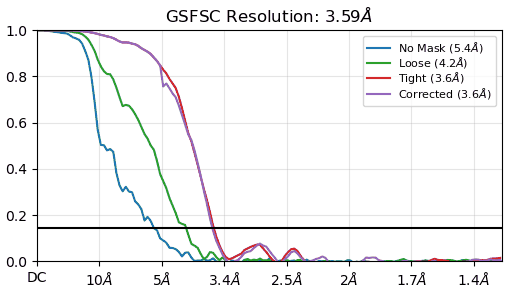} &
    \includegraphics[width=0.207\linewidth]{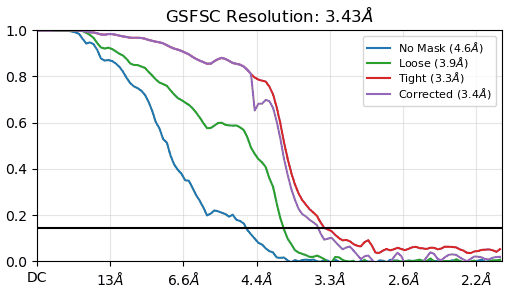} \\
    \parbox[b][0.120\linewidth][c]{2.0cm}{\footnotesize\raggedright Ours\\(fb)\\{\scriptsize full set}} &
    \includegraphics[width=0.207\linewidth]{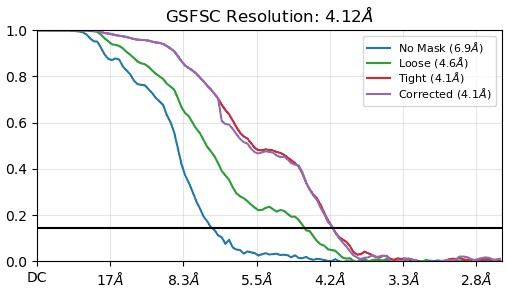} &
    \includegraphics[width=0.207\linewidth]{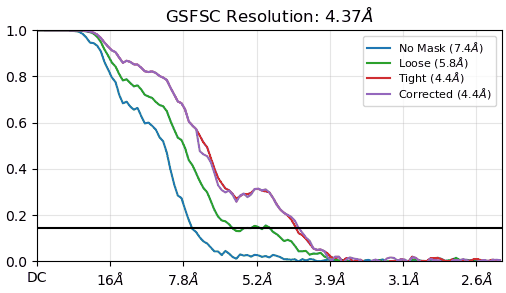} &
    \includegraphics[width=0.207\linewidth]{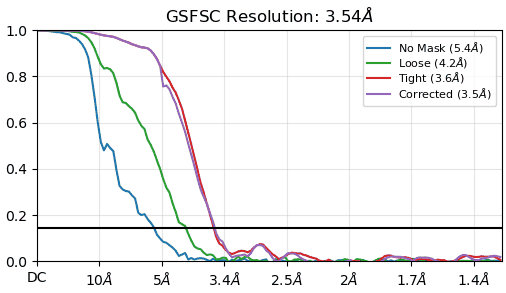} &
    \includegraphics[width=0.207\linewidth]{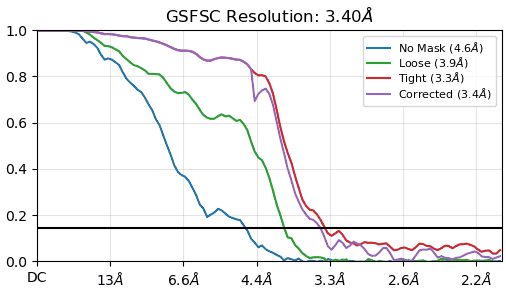} \\
    \multicolumn{5}{c}{\dotfill} \\
    \parbox[b][0.120\linewidth][c]{2.0cm}{\footnotesize\raggedright GT\\{\scriptsize 300 images}} &
    \includegraphics[width=0.207\linewidth]{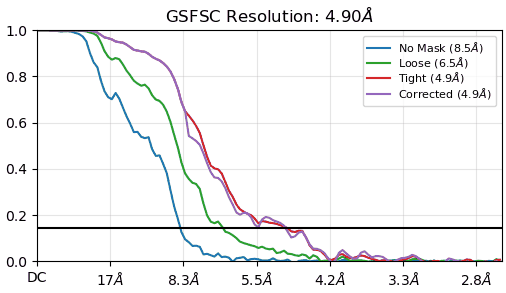} &
    \includegraphics[width=0.207\linewidth]{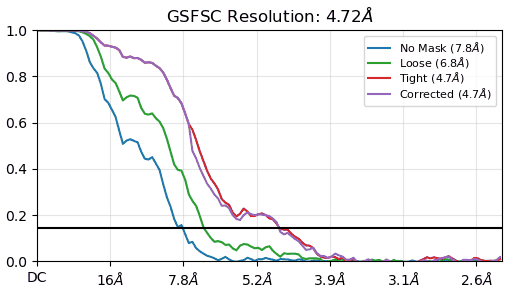} &
    \includegraphics[width=0.207\linewidth]{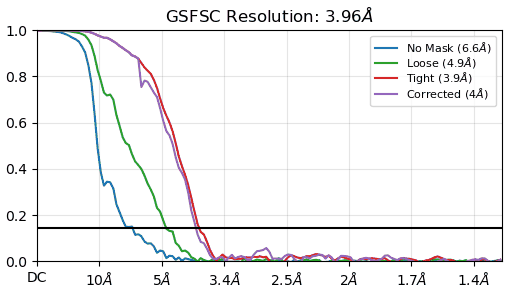} &
    \includegraphics[width=0.207\linewidth]{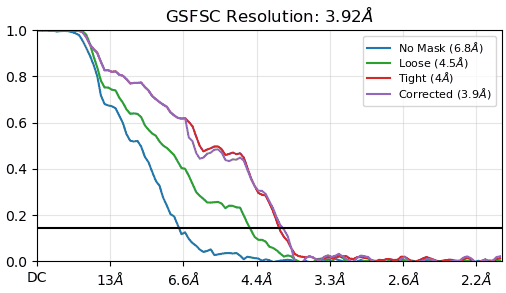} \\
  \end{tabular}
  \caption{\textbf{Gold-standard FSC of every reconstruction.} One column per dataset and
  one row per reconstruction, taken from the refinement that row delivers as its best of
  three seeds, with the resolution the main paper reports as the panel title. Below the
  dotted line is the reconstruction from the CryoPPP annotations of 300 micrographs per
  dataset. EMPIAR-10081 for crYOLO and every entry for Topaz are not held out.}
  \label{fig:diag_fsc}
\end{figure*}

\Cref{fig:diag_fsc,fig:diag_viewing} carry the CryoSPARC output behind the main results table and the ablation table of the main paper, one panel per dataset and row. The rows the two tables share are drawn once. The CryoTransformer row of the main results table is the baseline of the ablation table, and the Ours row is its fb. Both figures add the reconstruction from the CryoPPP annotations, which shows what the annotated particles themselves deliver. It is not one of the rows, since the annotations cover 300 micrographs of each dataset and every other row uses the full set.

In \cref{fig:diag_fsc}, the Corrected curve, which subtracts the correlation that the tight mask itself introduces, stays within 0.2\,\AA{} of the Tight curve on every panel, so no reported resolution rests on that correlation. The baseline and $+$mask rows of EMPIAR-10345 are the only panels where the four curves nearly coincide, and under $+$select the separation returns. Masking raises the FSC only where the half-maps agree beyond the solvent noise, so this coincidence indicates that little such signal is left, consistent with the precision of 0.187 of CryoTransformer on this entry (\cref{tab:f1_vs_res}). The GT row reaches a worse resolution than the Ours row on every entry, which we attribute to its particle count.

\Cref{fig:diag_viewing} plots the poses that each refinement assigns to its particles, so every panel is an estimated distribution rather than a ground-truth one, including the GT row, whose poses are also estimated. On EMPIAR-10081, the $+$both row has the most even distribution over directions. On EMPIAR-10345, the baseline and $+$mask rows populate nearly every direction and deliver 7.11 and 6.96\,\AA{} (\cref{fig:diag_fsc}). The rows after 2D selection deliver 3.54 to 3.59\,\AA{} from a few directions. The near-uniform coverage of the baseline row therefore does not come from genuine views. We attribute the strong bias that remains after the selection to the cycles discarding too many particles (\cref{sec:supp:selection}). On EMPIAR-10532, the $+$select, $+$both, and Ours rows occupy fewer directions than the baseline row while reaching better resolutions. The GT row shows that the annotated particles themselves are not uniformly distributed on any entry.

\section{2D detection scores of the base pickers}
\label{sec:supp:f1scores}

\Cref{tab:f1_vs_res} lists the scores behind the F1 plot of the main paper, measured against the CryoPPP annotations \cite{cryoppp2023}.

EMPIAR-10081 is part of the training data of crYOLO's general model, and the training data of the released Topaz general model is undocumented (the Topaz publication \cite{topaz2019} uses none of the four entries), so an overlap cannot be ruled out for Topaz. The affected values are grayed in \cref{tab:f1_vs_res} and excluded from its ranking.

These scores are also why the main paper builds on CryoTransformer. Among the pickers with no known training overlap with these entries, CryoTransformer has the highest recall on every entry, which suits a pipeline whose purification stages can only discard. CryoSegNet is trained on the same CryoPPP data and does not overlap the four entries either, which is why the main paper pairs it with CryoTransformer for the purification comparison.

\begin{table}[t]
  \centering
  \small
  \caption{\textbf{2D detection scores.} Macro precision, recall, and F1 on the 300
  annotated micrographs for the four base pickers. Gray marks the possible training
  overlap described in the text. Bold is best and underline second best per dataset
  among the black values.}
  \label{tab:f1_vs_res}
  \setlength{\tabcolsep}{3pt}
  \begin{tabular}{llcccc}
    \toprule
    Metric & Method & 10081 & 10093 & 10345 & 10532 \\
    \midrule
    \multirow{4}{*}{Prec.\ ($\uparrow$)}
      & crYOLO          & \textcolor{gray}{0.727} & \textbf{0.513} & \textbf{0.525} & \textbf{0.582} \\
      & Topaz           & \textcolor{gray}{0.476} & \textcolor{gray}{0.312} & \textcolor{gray}{0.330} & \textcolor{gray}{0.417} \\
      & CryoSegNet      & \textbf{0.664} & \underline{0.408} & \underline{0.415} & \underline{0.535} \\
      & CryoTransformer & \underline{0.469} & 0.352 & 0.187 & 0.459 \\
    \midrule
    \multirow{4}{*}{Rec.\ ($\uparrow$)}
      & crYOLO          & \textcolor{gray}{0.904} & \underline{0.422} & \underline{0.565} & \underline{0.305} \\
      & Topaz           & \textcolor{gray}{0.972} & \textcolor{gray}{0.908} & \textcolor{gray}{0.967} & \textcolor{gray}{0.839} \\
      & CryoSegNet      & \underline{0.821} & 0.340 & 0.543 & 0.194 \\
      & CryoTransformer & \textbf{0.954} & \textbf{0.737} & \textbf{0.966} & \textbf{0.620} \\
    \midrule
    \multirow{4}{*}{F1 ($\uparrow$)}
      & crYOLO          & \textcolor{gray}{0.801} & \underline{0.453} & \textbf{0.513} & \underline{0.375} \\
      & Topaz           & \textcolor{gray}{0.627} & \textcolor{gray}{0.463} & \textcolor{gray}{0.482} & \textcolor{gray}{0.549} \\
      & CryoSegNet      & \textbf{0.723} & 0.364 & \underline{0.451} & 0.270 \\
      & CryoTransformer & \underline{0.610} & \textbf{0.475} & 0.299 & \textbf{0.514} \\
    \bottomrule
  \end{tabular}
\end{table}

\section{Limitations of the evaluation}
\label{sec:supp:limitations}

Two limitations bound what these results establish. First, the 2D scores against the CryoPPP annotations are not held-out numbers, because 50 of the 300 annotated micrographs are also used for training in each round. Second, the resolution values on EMPIAR-10345 follow the pixel size declared in CryoPPP and are about half the physical figure, so they compare conditions within that entry only.

\twocolumn[{%
  \centering
  \setlength{\tabcolsep}{1.5pt}
  \renewcommand{\arraystretch}{0.9}
  \begin{tabular}{lcccc}
   & \footnotesize EMPIAR-10081 & \footnotesize EMPIAR-10093 & \footnotesize EMPIAR-10345 & \footnotesize EMPIAR-10532 \\
    \parbox[b][0.081\linewidth][c]{2.0cm}{\footnotesize\raggedright crYOLO\\{\scriptsize full set}} &
    \includegraphics[width=0.176\linewidth]{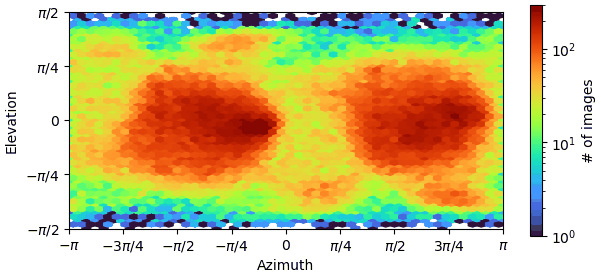} &
    \includegraphics[width=0.176\linewidth]{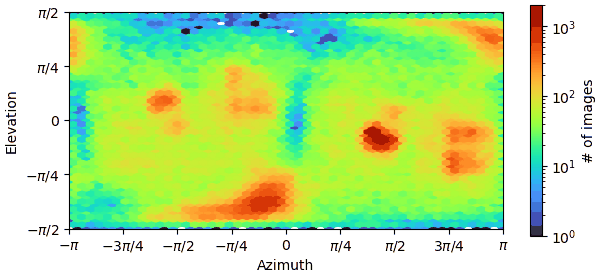} &
    \includegraphics[width=0.176\linewidth]{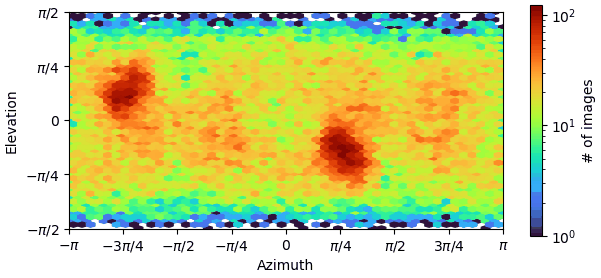} &
    \includegraphics[width=0.176\linewidth]{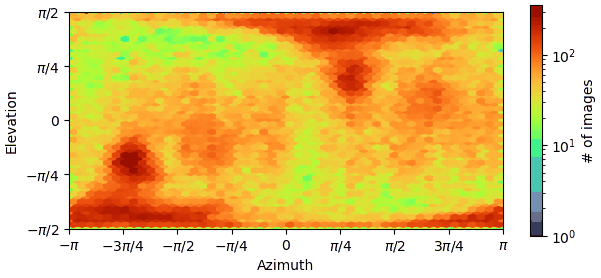} \\
    \parbox[b][0.081\linewidth][c]{2.0cm}{\footnotesize\raggedright Topaz\\{\scriptsize full set}} &
    \includegraphics[width=0.176\linewidth]{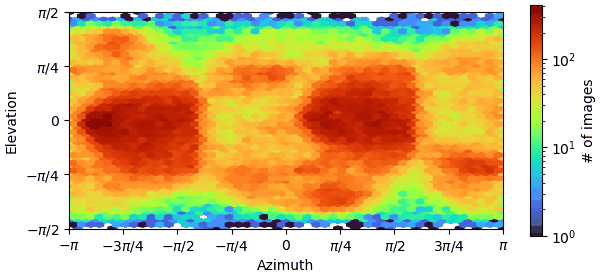} &
    \includegraphics[width=0.176\linewidth]{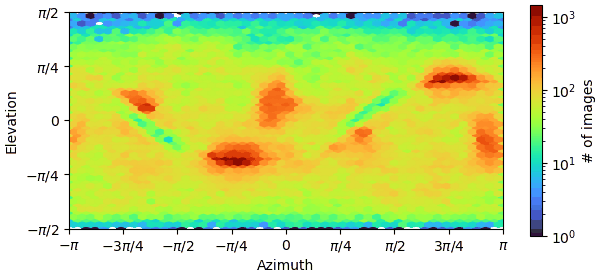} &
    \includegraphics[width=0.176\linewidth]{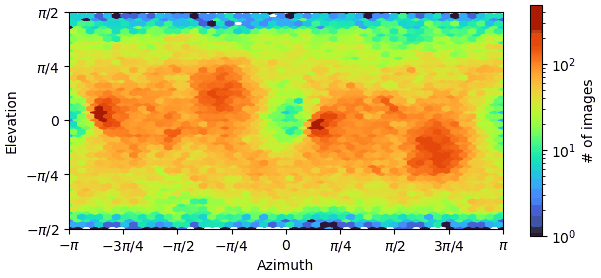} &
    \includegraphics[width=0.176\linewidth]{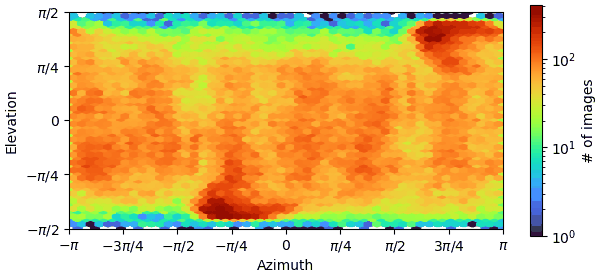} \\
    \parbox[b][0.081\linewidth][c]{2.0cm}{\footnotesize\raggedright CryoSegNet\\{\scriptsize full set}} &
    \includegraphics[width=0.176\linewidth]{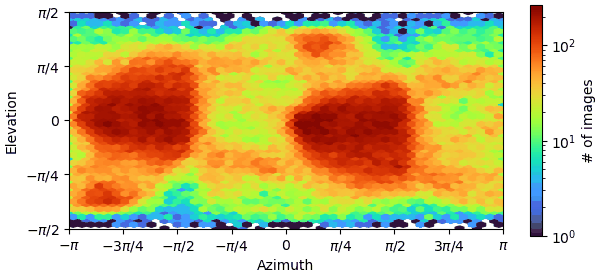} &
    \includegraphics[width=0.176\linewidth]{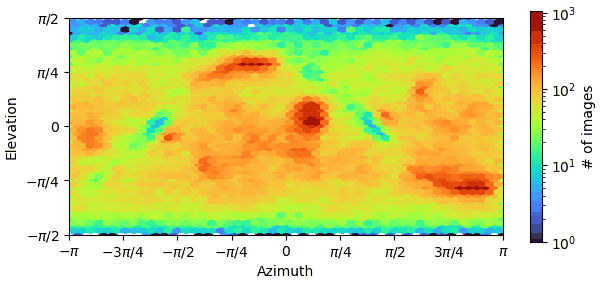} &
    \includegraphics[width=0.176\linewidth]{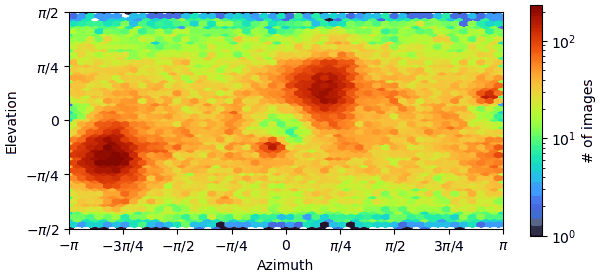} &
    \includegraphics[width=0.176\linewidth]{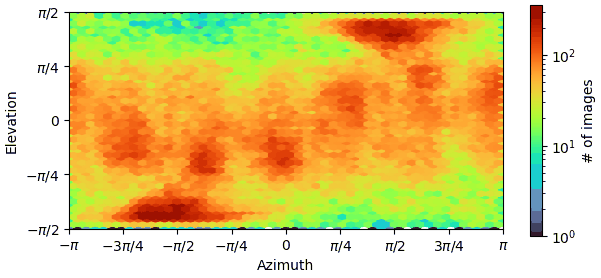} \\
    \parbox[b][0.081\linewidth][c]{2.0cm}{\footnotesize\raggedright CryoTransformer\\(baseline)\\{\scriptsize full set}} &
    \includegraphics[width=0.176\linewidth]{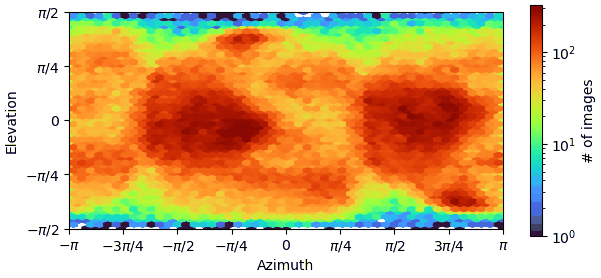} &
    \includegraphics[width=0.176\linewidth]{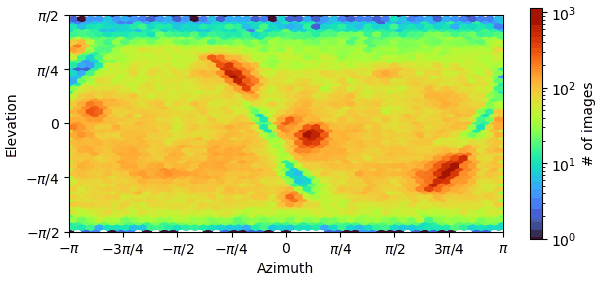} &
    \includegraphics[width=0.176\linewidth]{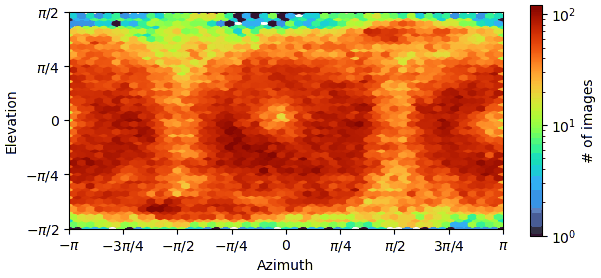} &
    \includegraphics[width=0.176\linewidth]{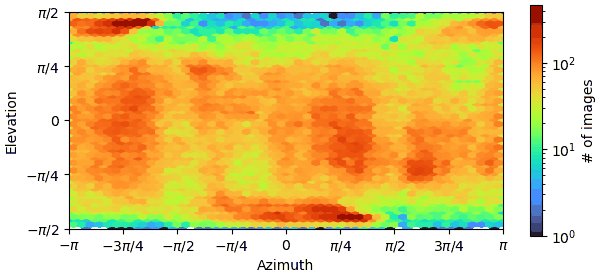} \\
    \parbox[b][0.081\linewidth][c]{2.0cm}{\footnotesize\raggedright $+$mask\\{\scriptsize full set}} &
    \includegraphics[width=0.176\linewidth]{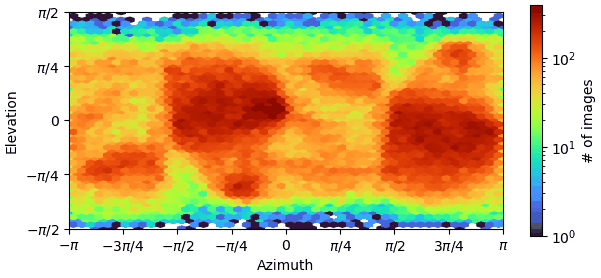} &
    \includegraphics[width=0.176\linewidth]{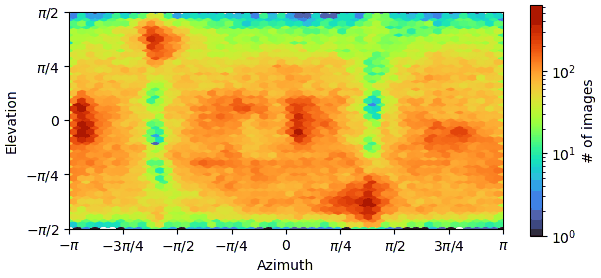} &
    \includegraphics[width=0.176\linewidth]{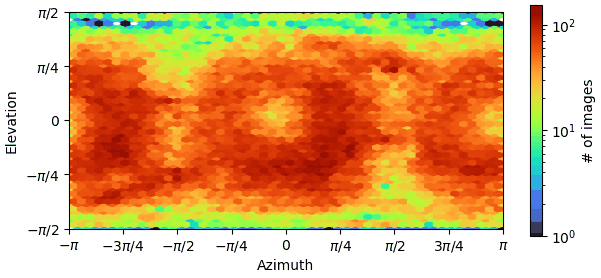} &
    \includegraphics[width=0.176\linewidth]{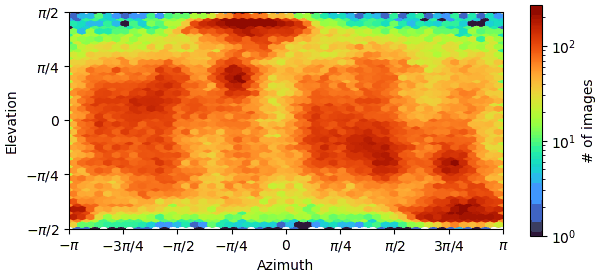} \\
    \parbox[b][0.081\linewidth][c]{2.0cm}{\footnotesize\raggedright $+$select\\{\scriptsize full set}} &
    \includegraphics[width=0.176\linewidth]{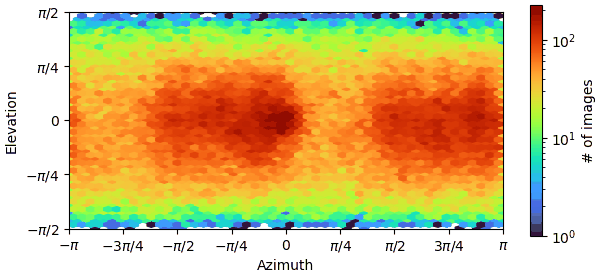} &
    \includegraphics[width=0.176\linewidth]{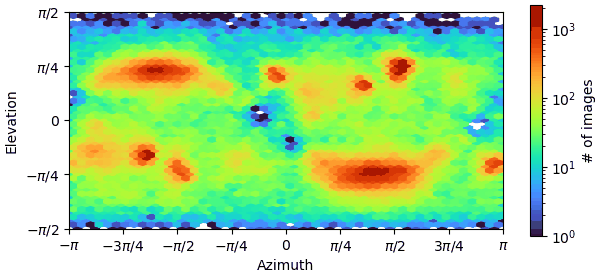} &
    \includegraphics[width=0.176\linewidth]{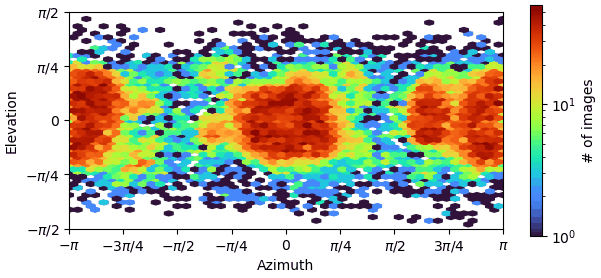} &
    \includegraphics[width=0.176\linewidth]{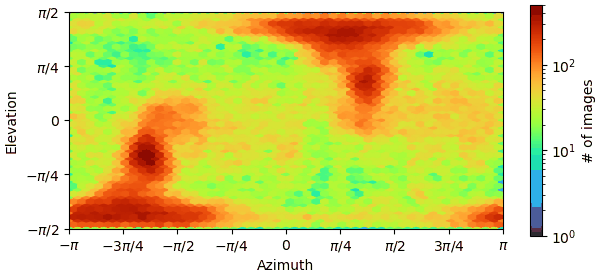} \\
    \parbox[b][0.081\linewidth][c]{2.0cm}{\footnotesize\raggedright $+$both\\{\scriptsize full set}} &
    \includegraphics[width=0.176\linewidth]{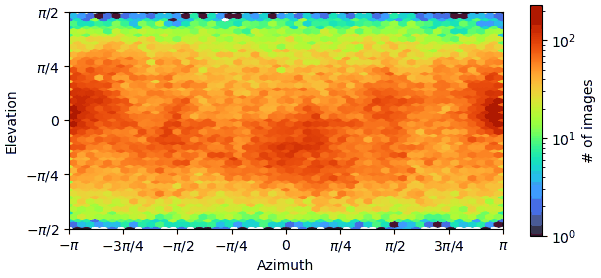} &
    \includegraphics[width=0.176\linewidth]{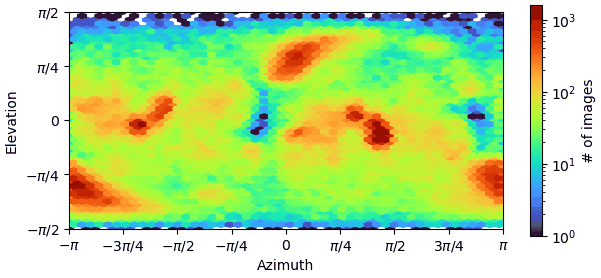} &
    \includegraphics[width=0.176\linewidth]{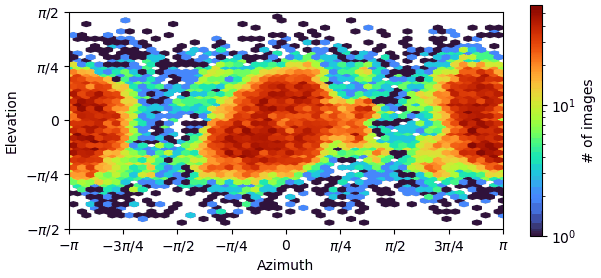} &
    \includegraphics[width=0.176\linewidth]{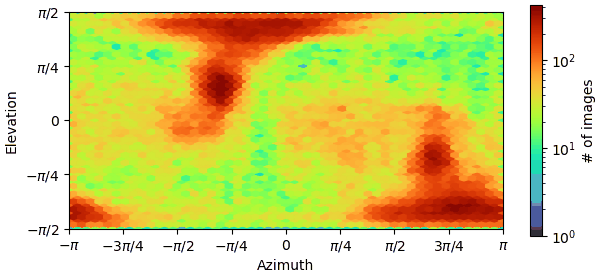} \\
    \parbox[b][0.081\linewidth][c]{2.0cm}{\footnotesize\raggedright Ours\\(fb)\\{\scriptsize full set}} &
    \includegraphics[width=0.176\linewidth]{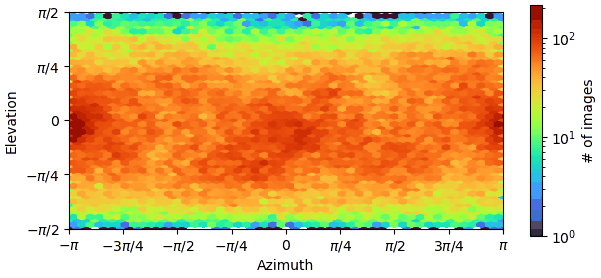} &
    \includegraphics[width=0.176\linewidth]{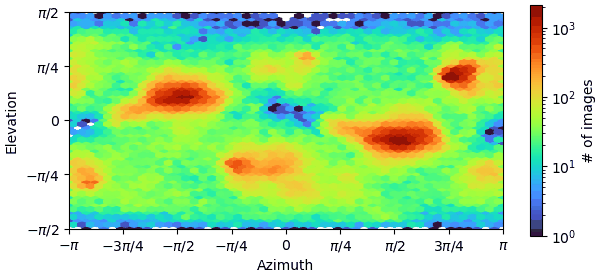} &
    \includegraphics[width=0.176\linewidth]{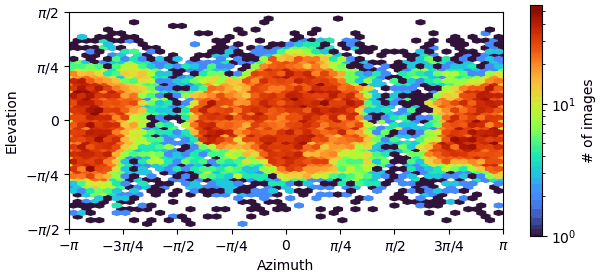} &
    \includegraphics[width=0.176\linewidth]{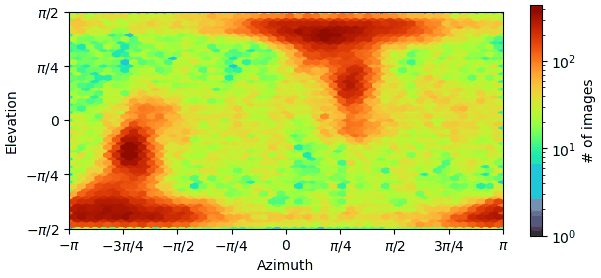} \\
    \multicolumn{5}{c}{\dotfill} \\
    \parbox[b][0.081\linewidth][c]{2.0cm}{\footnotesize\raggedright GT\\{\scriptsize 300 images}} &
    \includegraphics[width=0.176\linewidth]{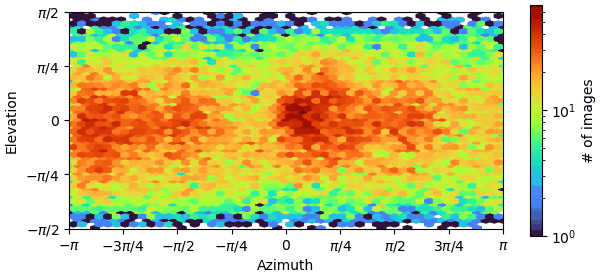} &
    \includegraphics[width=0.176\linewidth]{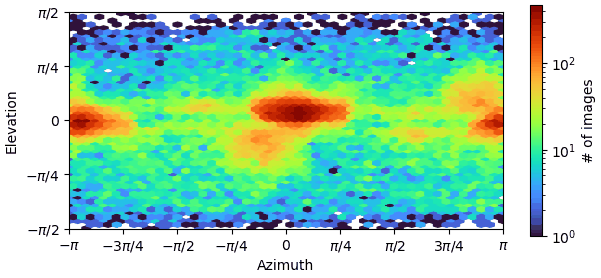} &
    \includegraphics[width=0.176\linewidth]{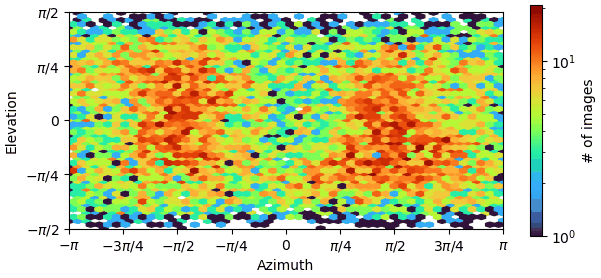} &
    \includegraphics[width=0.176\linewidth]{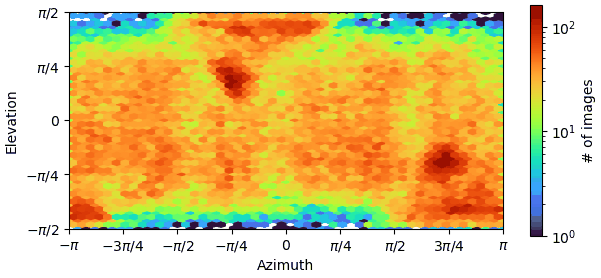} \\
  \end{tabular}
  \captionof{figure}{\textbf{Viewing directions of every reconstruction.} The distribution
  over azimuth and elevation of the same refinements as \cref{fig:diag_fsc}, in the same
  rows, where red is the most populated direction and blue the least. Below the dotted line
  is the reconstruction from the CryoPPP annotations of 300 micrographs per dataset.
  CryoSPARC scales each panel to its own counts, so the colors compare directions within a
  panel and not across panels.}
  \label{fig:diag_viewing}
  \vspace{16pt}
}]

\section{GPUs and compute time}
\label{sec:supp:cost}

Every experiment runs on one node with NVIDIA RTX A5000 GPUs of 24\,GB each, two AMD EPYC 7763 processors totaling 128 physical cores, and 2\,TB of RAM. Every stage of the pipeline occupies a single GPU. On the full micrograph sets, picking costs 0.49 to 0.55 seconds per micrograph, and applying the cached contamination masks to the resulting coordinates costs at most 75 seconds for an entire set. The cost sits in the stages that touch every particle. Extraction with 2D classification takes 2.6 to 5.0 hours, the CryoSift \cite{cryosift2025} cycles 1.3 to 8.5 hours, and one reconstruction arm of three ab-initio runs and three refinements 1.1 to 2.0 hours. A fine-tuning round costs the same on every dataset, since the teacher set is 50 micrographs, and its 50 epochs take just under two hours.

\end{document}